\documentstyle[12pt,epsf]{ioplppt}          
\newcommand{\eqb}{\begin{eqnarray}}
\newcommand{\eqe}{\end{eqnarray}}
\newcommand{\diff}{{\rm d}}
\newcommand{\adindex}{\hat{\gamma}}
\newcommand{\pomega}{{\bar\omega}}
\newcommand{\annalastro}[3]{{\it Annales d'Astrophys.}~{\bf#1}, #2}
\newcommand{\physicascripta}[3]{{\it Physica Scripta}~{\bf#1},#2}

\newcommand{\sovphysdokl}[3]{{\it Sov.\ Phys.\ Dokl.}~{\bf#1}, #2}
\newcommand{\sovphysjetp}[3]{{\it Sov.\ Phys.\ JETP}~{\bf#1}, #2}

\newcommand{\physrev}[3]{{\it Phys.\ Rev.}~{\bf #1}, #2}

\newcommand{\repprogphys}[3]{{\it Rep.\ Prog.\ Phys.}~{\bf #1}, #2}
\newcommand{\revmodphys}[3]{{\it Rev.\ Mod.\ Phys.}~{\bf #1}, #2}
\newcommand{\physrevletts}[3]{{\it Phys.\ Rev.\ Letts.}~{\bf #1}, #2}

\newcommand{\planetspacesci}[3]{{\it Planet.\ Space Sci.}~{\bf #1}, #2}

\newcommand{\apj}[3]{{\it Ap.~J.}~{\bf #1},~#2}

\newcommand{\physfluids}[3]{{\it Phys.\ Fluids}~{\bf #1},~#2}

\newcommand{\apjletts}[3]{{\it Ap.~J.~Letters}~{\bf #1},~#2}

\newcommand{\aanda}[3]{{\it
        As\-tron.\,As\-tro\-phys.}~{\bf #1}, #2}
\newcommand{\aandasuppl}[3]{{\it
        As\-tron.\,As\-tro\-phys.\ Suppl.}~{\bf #1}, #2}
\newcommand{\ass}[3]{{\it
        As\-trophys.\& Space Sci.}~{\bf #1}, #2}
\newcommand{\mnras}[3]{{\it M.~N.~R.~A.~S.}~{\bf #1}, #2}
\newcommand{\araa}[3]{{\it
       Ann.\,Rev.\,Astron.\,As\-tro\-phys.}~{\bf #1}, #2}

\newcommand{\nature}[3]{{\it Nature}~{\bf #1}, #2}
\newcommand{\jgr}[3]{{\it J.\,Geophys.\,Res.}~{\bf #1}, #2}

\newcommand{\jcompphys}[3]{{\it J.\,Comp.\,Phys.}~{\bf #1}, #2}
\newcommand{\jplasmaphys}[3]{{\it J.\,Plasma\,Phys.}~{\bf #1}, #2}
\newcommand{\commmathphys}[3]{{\it Commun.\,Math.\,Phys.}~{\bf #1}, #2}
\newcommand{\spscrev}[3]{{\it Space~Sc.~Rev.}~{\bf #1}, #2}
\newcommand{\icrcplodiv}[3]{{\it Proc.\ 15th.\ Int.\ Cosmic Ray Conf.}
                             (Plodiv) {\bf #1}, #2}

\begin{document}
\topical{Particle acceleration and relativistic shocks}

\author{J.G. Kirk\dag\ and P. Duffy\ddag}

\address{\dag\ Max-Planck-Institut f\"ur Kernphysik, Postfach 10 39 80, 
D--69029 Heidelberg, Germany}

\address{\ddag\ Department of Mathematical Physics, University College Dublin,
                Belfield, Dublin 4, Ireland} 

\begin{abstract}

Observations of both gamma-ray burst sources and certain classes of active
galaxy indicate the presence of relativistic shock waves and require the 
production of high energy particles to explain their emission. In this 
paper we first 
review the basic theory of shock waves in relativistic
hydrodynamics and magneto-hydrodynamics, emphasising the astrophysically interesting cases.
This is followed by an overview of the theory of particle acceleration at such shocks.
Whereas, for diffusive acceleration at 
non-relativistic shocks, it is the compression ratio which fixes the energetic
particle spectrum uniquely, acceleration at relativistic shocks is more 
complicated. In the absence of scattering, particles are simply 
\lq compressed\rq\ 
as they pass through the shock front. This mechanism -- called shock-drift 
acceleration -- enhances the energy density in accelerated particles, but does so 
without changing the spectral index of upstream particles. Scattering due
to MHD waves leads to multiple encounters between the particles and the shock 
front, producing an energetic particle
population which depends on the properties of the shock front and the level and 
nature of particle scattering. We describe the method of matching the angular
distributions of the upstream and downstream distributions at the shock front 
which leads to predictions of the spectral index. Numerical simulation of particle transport 
provides an alternative means of calculating spectral indices, and has recently been extended to 
cover ultra relativistic shocks. We review these calculations and summarise the applications to 
the astrophysics of relativistic jets and fireball models of gamma-ray-bursts.

\end{abstract}

%
%
\section{Introduction}
Within the space of thirty minutes, the power emitted in very high energy 
gamma-rays by an active galaxy at a 
distance of 
several hundred million 
light years has been observed to increase more than twenty fold \cite{gaidosetal96}. 
This unexpected observation 
has raised new questions concerning the nature of \lq gamma-ray blazars\rq, as 
this class of source is called. 
In a parallel development, the discovery that at least some of the 
enigmatic, powerful so-called \lq gamma-ray burst\rq\ sources lie at cosmological distances 
\cite{metzgeretal97} was perhaps less surprising to many astrophysicists, but nonetheless places stringent 
restrictions on the conditions prevalent in the emission regions. 
These two recent developments have been responsible for a renewed focusing of 
attention on the subject of relativistic shock fronts. 
In each case, the observations imply that the photon density in the source is so large, that the gamma-rays would 
be subject to strong absorption due to photon-photon interactions, were not the source moving in our direction 
with a speed close to that of light. The implied Lorentz factors are about 50-100 
in the case of the active galaxies, and several hundreds in the case of gamma-ray burst sources.

The subject is not new. Relativistic motion was proposed as a way out of the
rapid variability 
problem over thirty years ago \cite{rees66}. 
The observational motivation then was similar to that of today: in order for the 
variable synchrotron emission in the radio band observed from active galaxies to emerge without being subject to 
self-absorption an improbably weak magnetic field was implied. The proposed solution was that the 
source moves at relativistic speed towards the observer, a Lorentz factor of 5--10 being sufficient to remove the 
difficulty. The hypothesis was impressively confirmed by sequences of radio observations of 
the cores of active galaxies, using very long baseline radio interferometry, which clearly showed rapid expansion. 
This apparently \lq superluminal\rq\ 
motion is now seen in many sources, and indicates the outflow of plasma
containing relativistic electrons 
and magnetic fields with bulk 
Lorentz factors between roughly 5 and 20 \cite{vermeulencohen94}.  

Shock fronts are an inevitable consequence when such flows encounter the ambient material in the interstellar 
medium of the host galaxy, or in the intergalactic medium. Indeed, they seem
the natural sites for the 
dissipation 
of the kinetic energy associated with the relativistic motion itself. 
Furthermore, wherever we directly observe shock fronts in astrophysics -- from the Earth's bow shock to 
supernova remnants -- they are associated with particle acceleration. 
The radiation from both gamma-ray bursts 
and gamma-ray blazars requires the presence of particles (most likely electrons and/or positrons) with 
very high Lorentz 
factors (exceeding $10^6$ in blazars), so that particle acceleration must be happening there too.
The aim of this review is to summarise our current theoretical picture of 
relativistic shocks in astrophysics and, in particular, our ideas about the way in which they might accelerate 
particles.
Section~\ref{shock} presents an overview of hydrodynamic and MHD
shocks, Section~\ref{acceleration} examines shock drift acceleration as well as 
analytic and numerical methods of treating Fermi acceleration and
Section~\ref{discussion} briefly summarises the applications 
to relativistic jets and gamma-ray bursts.


\section{Relativistic Shocks}
\label{shock}
Relativistic effects can be important at shock fronts for two
distinct reasons.
Firstly, the post shock temperature can be so high that the
thermal motion of individual particles approaches the speed of
light. Secondly, it is possible that so much energy is released
into a region containing relatively little matter that
the subsequent expansion proceeds at a bulk speed
approaching that of light. It is certainly possible for the former
effect to arise when all bulk velocities are quite modest. For example,
a hot plasma
in which the pressure is provided predominantly by photons is always
relativistic in this sense, and behaves as a gas whose ratio of
specific heats is 4/3.
If the mass density in such a plasma is dominated by atoms or ions,
there exists a sound velocity which is determined by the pressure of
the photons and the inertia of the ions and which can be small
compared to the speed of light. Downstream of
a strong shock front in such a plasma,
the bulk speed is, of course, less than this sound
speed, whereas the upstream fluid speed (provided it is still
non-relativistic) is
equal to seven times the downstream speed.
On the other hand, it seems unlikely that in a realistic situation
a relativistic
bulk flow could be thermalised by a shock front without producing thermal
velocities which are also relativistic.
As a consequence, the fully relativistic
equation of state of a plasma should be used to investigate the
possible jump conditions.

Following the solution of the relativistic shock problem in hydrodynamics by
Taub~\cite{taub}, the question of which types of shock
discontinuities are permitted and what the jump conditions are in the  
ideal MHD picture was investigated by de~Hoffmann \& 
Teller \cite{dehoffmannteller50}. The full relativistic theory 
was presented by Akhiezer \& Polovin
\cite{akhiezerpolovin59} and an elegant summary of the most important results
has been given by Lichnerowicz~\cite{lichnerowicz70}. 
However, the generality of these treatments makes them less accessible to the
plasma astrophysics community, 
and several papers have  
worked out special cases, sometimes using  
particular approximations to the equation of state
\cite{webbetal87,majoranaanile87,applcamenzind88}.

In view of this, we present in this Section the basic, 
astrophysically relevant, 
results on both hydrodynamic and MHD shocks in as simple a form as is
compatible with completeness. In each case a numerical treatment is 
in general needed to
solve for the jump conditions, and we provide in the Appendices a detailed
description of how such an algorithm can be constructed. 
In the case of hydrodynamic shocks, the approximate formulae for 
strong shocks and for the ultra-relativistic case are presented, their
derivations being relegated to the Appendix. The special cases which are of
most interest in MHD are those of weak magnetic field and of ultra
relativistic, perpendicular shocks. Again, the relevant results are quoted and
the derivations presented in the Appendix. We also note that ultra-relativistic
switch-on shocks are possible only in a very limited parameter range.

The most important property of a shock front for the acceleration
problem is its compression ratio, and it has been known for some time
\cite{kundtkrotscheck82} that increasing the magnetic field strength tends to reduce the
compression. We present some figures at the end of the section which illustrate
this effect.

\subsection{Hydrodynamics}
It is useful at this point to recall the theory of
relativistic hydrodynamics and the shock wave solutions it admits. This will
also serve as a means of introducing the notation we will use throughout this
paper. 

In the absence of external forces and energy sources, 
the equations of relativistic 
hydrodynamics can be formulated as the vanishing 
divergence of the stress-energy tensor
associated with the fluid:
\eqb
\nabla_\mu T^{\mu\nu}&=&0
\label{fluideqs}
\eqe
(e.g., \cite{landaulifshitz59}). 
Neglecting dissipative effects, 
the stress-energy tensor is diagonal in the local plasma rest frame, and is 
given by
\eqb
T^{\mu\nu}&=&w u^\mu u^\nu + p g^{\mu\nu}
\eqe
Here $u^\mu$ is the four velocity of the fluid 
($\mu=0,1,2,3$), 
and $g^{\mu\nu}$ the metric
tensor, for which we adopt the convention $-$$+$$+$$+$. The scalars $w$ and $p$ are the
proper enthalpy density and pressure, i.e., those 
measured in the rest frame of the fluid, in which $u^\mu=(1,0,0,0)$. 

The problem of solving the Rankine-Hugoniot relations to 
find the jump conditions across a shock front requires one to use an equation
of state to find the quantity
$p/\rho$, given the quantity $e/\rho$. For a dissipation free ideal gas, the
Synge equation described in \ref{hydrojump} is appropriate, and it is
necessary to solve
Eq.~(\ref{eqofstate}) numerically. Although this procedure is
straightforward, a popular short-cut is to define 
a parameter ${\adindex}$ via the equation
\eqb
p&=&(\adindex-1)(e-\rho)
\label{fixedgamma}
\eqe
\cite{blandfordmckee76,koenigl80},
which replaces the equation of state (\ref{eqofstate}). In the non-relativistic
case ${\adindex}=5/3$ and can be identified as the ratio of specific heats
of the gas.
For a gas whose pressure is dominated by a relativistic component, one has
${\adindex}=4/3$ (a fully relativistic gas has, in addition, $e\gg\rho$).
For a gas consisting of equal numbers of 
electrons and protons in thermodynamic equilibrium with each other, there is
a range of temperatures $m_{\rm p}\gg T \gg m_{\rm e}$ 
in which $\adindex\approx13/9$
(here $m_{\rm e}$ is the electron mass and $m_{\rm p}$ the proton mass).

The jump conditions across a relativistic hydrodynamic shock are given by
Eq.~(\ref{fluideqs}), which expresses the laws of energy and momentum
conservation, together with the law of conservation of particle number. 
For those 
constituents which are neither created nor annihilated in the shock, one has
\eqb
\nabla_\mu (n_i u^\mu)&=&0
\label{particleconsv}
\eqe
The shock wave is a surface in space-time, 
$\phi(x_\mu)=0$, across which there is a discontinuity in
the fluid variables. 
The shock normal is given by $\l_\mu=\partial_\mu\phi$ and, without 
loss of generality, we normalise $\phi$ so that $l^\mu l_\mu=1$. 
Two frames of reference which are important from a physical point of view
are those in which the upstream and downstream plasma is at rest.
Defining the Lorentz scalars $v_-$ and $v_+$ to be 
the shock speed measured in the upstream and
downstream rest frames respectively, one has
\eqb
v_{\pm}&=&{{|u_\pm^\mu l_\mu|}\over{\sqrt{1+{|u_\pm^\mu l_\mu|}^2}}}.
\eqe
where  $u^\mu_-$ and $u^\mu_+$ are the 4-velocities of the plasma immediately
upstream and immediately downstream of the shock front, respectively.

Whenever necessary, 
we will use Cartesian coordinates in which the shock normal is
along the $x$-axis. In the upstream and downstream rest frames
$\l_\mu=(\Gamma_\pm v_\pm,\Gamma_\pm,0,0)$, where 
$\Gamma_\pm=(1-v_\pm^2)^{-1/2}$. 
As we show below, 
for a hydrodynamic shock front, the downstream plasma flows along the shock
normal, as seen from the upstream rest frame, so that in this frame 
$u_+^\mu=(\Gamma_{\rm rel},-\Gamma_{\rm rel}v_{\rm rel},0,0)$, where 
\eqb
v_{\rm rel}&=& {v_- - v_+\over 1 - v_-v_+}
\label{relvel}
\eqe
is the relative speed of the upstream gas with respect to the downstream gas,
and its related Lorentz factor is $\Gamma_{\rm rel}$.
 
From the conservation laws, the shock jump conditions are 
\eqb
\left[n_i u^\mu\right]l_\mu&=&0
\label{hydrojumpcond1}\\
\left[T^{\mu\nu}\right]l_\mu&=&0.
\label{hydrojumpcond2}
\eqe
where (\ref{hydrojumpcond1}) is valid for all conserved constituents.
For non-conserved particles additional 
information is required specifying the 
number density, for example via the equation
of state. In the shock rest frame, with the shock normal and upstream flow
velocity along the $x$-axis it is straightforward to show that the downstream
flow velocity is also directed along the same axis. In this frame 
equations (\ref{hydrojumpcond1}) and (\ref{hydrojumpcond2}) can be written as
\eqb
\Gamma_-\rho_-v_-&=&\Gamma_+\rho_+v_+
\label{hydrojumpsimp1}\\
\Gamma_-^2w_-v_-^2+p_-&=&\Gamma_+^2w_+v_+^2+p_+
\label{hydrojumpsimp2}\\
\Gamma_-^2w_-v_-&=&\Gamma_+^2w_+v_+
\label{hydrojumpsimp3}
\eqe
Given $v_-$ and the upstream state, $e_-/\rho_-$, these equations 
are to be solved for the downstream  state $v_+$, $e_+/\rho_-$ and
the proper compression ratio $R\equiv\rho_+/\rho_-$.
In general, this entails a numerical
procedure, which is described in \ref{hydrojump}. However, 
there are several interesting special cases with analytic solutions:

\begin{enumerate}
\item
In the limit of non-relativistic fluid speeds, $v_\pm\ll1$, the
well-known Rankine-Hugoniot conditions are recovered. 
In terms of the upstream Mach number 
\eqb
M_-&=&v_-\left({{\adindex}p_-\over\rho_-}\right)^{-1/2}
\label{machnumber}
\eqe
one has
\eqb
p_+&=&\rho_- v_-^2\left[{2\over(\adindex+1)}- {(\adindex-1)\over\adindex
(\adindex+1)M_-^2}\right]
\nonumber\\
v_+&=&v_-\left[{(\adindex-1)\over(\adindex+1)}+{2\over(\adindex+1)M_-^2}
\right]
\nonumber\\
{\rho_+\over\rho_-}&=&{\adindex+1\over\adindex-1+(2/M_-^2)}
\eqe
where in our case of a monatomic gas, the relevant value of the adiabatic
index is $\adindex=5/3$.
\item
For a shock in a relativistic gas, in which $p=e/3$ (both upstream and
downstream) one finds the simple relation
\eqb
v_- v_+={1\over3}
\label{simplejump}
\eqe
However, this corresponds to the perhaps less interesting case of a relativistic
shock moving into unshocked gas in which the total energy density significantly
exceeds that corresponding to rest-mass i.e., $e_-\gg \rho_-$.
\item
For a strong shock, at which the 
upstream pressure can be neglected,
the equation of state in the upstream gas is unimportant.
Furthermore, if particles are conserved at the shock, 
the average Lorentz factor of a particle does not change
across the shock as seen from the downstream rest frame, i.e., 
$e_+=\Gamma_{\rm rel}\rho_+$ (see~\ref{hydrojump}, 
(Eq.~\ref{consavlor}) with $\eta=1$). 
Then, using, for the downstream medium, 
the equation of state for fixed adiabatic index (Eq.~\ref{fixedgamma})
one may derive
\eqb
w_+/\rho_+&=&{\adindex}(\Gamma_{\rm rel}-1)+1
\nonumber\\
\Gamma_-^2&=&{ (w_+/\rho_+)^2(\Gamma_{\rm rel}+1)\over
{\adindex}(2-{\adindex})(\Gamma_{\rm rel}-1)+2}
\label{fixbland}
\eqe
\cite{blandfordmckee76},
which gives the shock
speed and downstream pressure in terms of the relative velocity of the upstream
and downstream fluids. These relations are generalised in \ref{hydrojump}
to the case in which particles are 
created at the shock front.
\item
In the ultra-relativistic limit, $\Gamma_-\rightarrow\infty$, 
the upstream pressure ($p_-$) may be neglected in 
Eq. (\ref{hydrojumpsimp2}). If, in addition, the downstream particles are ultra-relativistic, in the sense that 
$e_+\gg\rho_+$, one may combine Eqs.~(\ref{hydrojumpsimp2}),
(\ref{hydrojumpsimp3}) and (\ref{relvel}) 
to find:
\eqb
v_+&\rightarrow&\adindex-1\,=\,1/3
\\
\Gamma_{{\rm rel}}&\rightarrow&\Gamma_-\sqrt{(2-\adindex)/\adindex}
\,=\, \Gamma_-/\sqrt{2}
\eqe
These relations are independent 
of the equations of state upstream and downstream and hold whether or not
particles are conserved at the shock, 
provided only that the downstream particles are 
ultra-relativistic.
\end{enumerate}

\subsection{Magnetohydrodynamic Shocks}

In ideal, relativistic MHD, it is assumed that the plasma is dissipation free
and that in the local rest frame the electric field vanishes. 
In this case, the
electromagnetic field is specified by the magnetic field alone, 
and the system is described by
the four \lq source-free\rq\ equations of Maxwell:
$\nabla_\mu(*F^{\mu\nu})=0$ where $*F_{\mu\nu}$ 
is the dual electromagnetic field tensor,
\eqb
*F_{\mu\nu}=
\left(\matrix{0&B_1&B_2&B_3\cr
-B_1&0&E_3&-E_2\cr
-B_2&-E_3&0&E_1\cr
-B_3&E_2&-E_1&0\cr}\right)
\eqe
where ${\bf B}=(B_1,B_2,B_3)$ and ${\bf E}$
are the magnetic and electric fields
as measured in a frame where the plasma's three velocity is ${\bf\vec{\beta}}$, augmented by the generalised 
(ideal) Ohm's Law ${\bf E}
=-{\bf\vec{\beta}\times B}$, which constrains the electric field to vanish in the plasma rest frame.
Defining the four-vector
$B_\mu=-u^\nu(*F_{\mu\nu})$ the field tensor can be written 
as $*F^{\mu\nu}=B^\mu u^\nu-u^\mu B^\nu$
so that the source-free Maxwell equations become
\eqb
\nabla_\mu\left(B^\mu u^\nu-u^\mu B^\nu\right)=0.
\label{maxwell}
\eqe
In the rest frame of the plasma $B^\mu=(0,{\bf B})$ where ${\bf B}$ is 
the magnetic field three vector in that frame. In the following, $B$ is 
taken to denote the magnetic field strength in the local plasma rest 
frame which satisfies $B^\mu B_\mu=B^2$. The components of $B^\mu$ in a 
frame where the plasma is moving with four velocity $u^{\mu}$ can be 
derived from the appropriate Lorentz transformation in terms of 
${\bf B}$. 
The energy momentum tensor of the system consisting of electromagnetic fields and fluid is
\eqb
T^{\mu\nu}&=&\left(w+{B^2\over 4\pi}\right)u^\mu u^\nu +
\left(p+{B^2\over 8\pi}\right)g^{\mu\nu} - {{B^\mu B^\nu}\over 4\pi}.
\label{mhdstress}
\eqe
and the equations of relativistic MHD consist of the vanishing divergence of this tensor, together with 
(\ref{maxwell}) and, of course, an equation of state for the fluid. 
An important difference between relativistic and non-relativistic MHD is that 
in relativistic MHD it is not possible simplify (\ref{maxwell}) by ignoring 
the displacement current. This is because in the relativistic case, fluctuations in the space charge cannot be 
neutralised on a time scale which is much faster than the other dynamical 
time-scales.

There are three wave modes in MHD, the fast and slow magnetosonic modes, and
the Alfv\'en mode. The phase velocities of these modes in a given direction 
will be
denoted by $v_{{\rm fast}}$, $v_{{\rm slow}}$ and $v_{{\rm A}}$. 
Quite generally one has
\eqb
v_{{\rm fast}}\ge v_{{\rm A}}\ge v_{{\rm slow}} 
\eqe
Defining $\Phi$ as the angle between the magnetic field and the direction of
propagation, measured in the plasma rest frame, the Alfv\'en wave has the speed
\eqb
v_{{\rm A}}&=&
\sqrt{B^2/(4\pi)\over w+B^2/(4\pi)} \cos\Phi
\label{alfvenspeed}
\eqe
and the fast and slow mode speeds are the roots of the equation
\eqb
w(v_{\rm s}^{-2}-1)v^4-\left[w+{B^2\over 4\pi v_{\rm s}^{2}}\right]v^2(1-v^2)
+{B^2\over 4\pi}\cos^2\Phi(1-v^2)&=&0
\label{fastslowspeed}
\eqe
\cite{majoranaanile87}.
These wave speeds are important for the
properties of shock fronts 
\cite{akhiezerpolovin59,lichnerowicz70}, 
since, as in the non-relativistic case
\cite{akhiezeretal59,kenneletal83}, 
a physically realisable shock front cannot provide a transition
across the Alfv\'enic point in ideal MHD. 
In other words, the two kinds of transition 
available are the slow-mode shock, with
\eqb
v_{{\rm A}-} > v_- > v_{{\rm slow}-}
\nonumber\\
v_{{\rm A}+} > v_{{\rm slow}+} > v_+
\label{evolutionarity1}
\eqe
and the fast-mode shock which has
\eqb
v_- > v_{{\rm fast}-} > v_{{\rm A}-} 
\nonumber\\
v_{{\rm fast}+} > v_+ > v_{{\rm A}+}
\label{evolutionarity2}
\eqe
where $v_{{\rm fast,A,slow}\pm}$ refer to the propagation speeds along the 
normal to the shock front in the up and downstream regions.
The fast-mode shock is associated with an increase in the magnetic field
strength across the shock, whereas the slow-mode shock results in a decrease of
the field.

For the purposes of numerical simulation of a relativistic MHD flow, it 
is interesting to 
put the governing equations 
into divergence form, avoiding the additional
algebraic constraint implied by the relation $u^\mu B_\mu=0$. 
Such a formulation has been presented \cite{vanputten91} and 
implemented \cite{vanputten93} by
van~Putten. 
However, at least for the non-relativistic case, 
other methods are available
(e.g., Falle et al.~\cite{falleetal98}) and there is also
no advantage to 
be gained by such a reformulation in the investigation of
the jump conditions. These
are determined by the equations of conservation of
particles and of energy/momentum (\ref{hydrojumpcond1}) and
(\ref{hydrojumpcond2}) [as in the hydrodynamic case, but with the
modified stress-energy tensor of Eq.~(\ref{mhdstress})], supplemented 
by  equation~(\ref{maxwell}) which implies:
\eqb
\left[B^\mu u^\nu-u^\mu B^\nu\right]l_\mu=0.
\label{mhdjumpcond3}
\eqe 
It proves convenient to define $b^\mu_\pm\equiv B^{\mu}_\pm/
\sqrt{4\pi\rho_-}$ so that $|b|^2_\pm=B^2_\pm/4\pi\rho_-$ is {\em twice} 
the ratio of the magnetic field energy 
density in the local plasma rest frame to the upstream rest mass energy density. 
In the upstream and downstream rest frames we define $\Phi_\pm$ to be the angle
between the magnetic field and the shock 
front's direction of propagation, and take coordinates such that the field lies
in the $x$--$z$ plane. Therefore we can write for the two vectors 
$b_+$ and $b_-$
\eqb
b^\mu_\pm&=&(4\pi\rho_-)^{-1/2}(0,B_\pm\cos\Phi_\pm,0,B_\pm\sin\Phi_\pm)
\label{defsmallb}
\eqe
in
the upstream or downstream frame as appropriate.
Given $v_-$, $\Phi_-$ and $|b|_-^2$, 
the jump conditions can be solved numerically to find $v_+$, $\Phi_+$, $|b|_+^2$ and the
compression ratio $\rho_+/\rho_-$
using the method of
Majorana \& Anile \cite{majoranaanile87}, which is described in
\ref{mhdjump}, and 
involves a single one-dimensional root-finding operation. 

In general, the plane which contains the shock normal and the upstream magnetic
field, as seen in 
the upstream rest frame, also contains the plasma velocity and the 
magnetic field in the downstream half-space -- i.e., these shocks do not permit
the generation of a non-coplanar magnetic field. However, in addition to the
fast-mode and slow-mode shocks, there exist solutions of the jump conditions
(known as rotational discontinuities), which propagate with the Alfv\'en speed
and do not involve a compression of the plasma, or a change in the field
strength, but merely a change in the direction of the field and the plasma
speed. In the standard picture of Fermi acceleration, they are not expected to
accelerate particles. Slow mode shocks are also thought to be ineffective
accelerators. Not only does the magnitude of the magnetic field decrease across
the shock, (rendering shock drift acceleration ineffective)
but also, if particle scattering is produced by Alfv\'en waves, then the
speed of the scattering centres in the fluid frame exceeds the speed of the
slow mode shock. Crossing and recrossing would then lead to a net deceleration
if the waves predominantly stream away from the shock front, which would
turn diffusive acceleration into diffusive deceleration \cite{isenberg86}. 
Therefore, only
fast-mode shocks are considered in the following.

As in the hydrodynamical case, there are some interesting limiting cases for 
MHD shocks:
\begin{enumerate}
\item In the case of a weak, dynamically unimportant magnetic field, such as
considered by Kirk \& Heavens~\cite{kirkheavens89}, Begelman \&
Kirk~\cite{begelmankirk90} several useful
analytic results can be obtained.  Of course, the jump conditions for the
plasma speed and temperature are identical to the hydrodynamical case. In the 
shock rest frame with both $v_-$ and $v_+$ directed along the $x$-axis, the 
$x$-component of magnetic field is conserved across the shock. Since the component
of magnetic field in the direction of the flow velocity is unchanged by a Lorentz
boost to the local fluid frame this gives a simple expression connecting the angles
$\Phi_-$ and $\Phi_+$,
\eqb
B_-\cos\Phi_-=B_+\cos\Phi_+
\label{bparformula}
\eqe
The (proper) 
compression in the magnetic field can be related to the upstream parameters
and the compression ratio $r=v_-/v_+=R\Gamma_+/\Gamma_-$ of the fluid:
\eqb
R_B\equiv {B_+\over B_-}&=&
\left[r^2 - \Gamma_-^2(r^2-1)(\cos^2\Phi_- - v_-^2)\right]^{1/2}
\label{khformula}
\eqe
\item
If the upstream plasma flows along the magnetic field and the shock normal, 
it is possible to 
have a purely hydrodynamic transition in which the downstream flow is also
along the magnetic field. However, for some parameter ranges, it is also 
possible to generate a transverse component of the field at the shock front in
what is called a \lq switch-on\rq\ shock. 
This happens if the hydrodynamic
solution violates the 
constraints given by (\ref{evolutionarity1}) and  
(\ref{evolutionarity2}) (referred to as  evolutionarity
conditions\footnote{Lichnerowicz~\protect\cite{lichnerowicz67} 
has pointed out that
also the switch-on (and the switch-off) shock is not evolutionary. The
physical reason for this is that the plasma cannot break the azimuthal symmetry
around the direction of the magnetic field in order to create the transverse
component. However, for an arbitrarily small angle between the shock normal and
magnetic field, this symmetry is broken, and a shock appears which is both
evolutionary and arbitrarily \lq close\rq\ to the switch-on shock. 
On the other hand, the purely hydrodynamic solution does not have an
evolutionary shock in its neighbourhood, when conditions (\protect\ref{evolutionarity1})
and (\protect\ref{evolutionarity2}) are violated \protect\cite{majoranaanile87}.}). 
To find the relevant range of parameters 
in the ultra-relativistic case, we make the approximations
\eqb
v_-\rightarrow1;\qquad v_+\rightarrow1/3;\qquad \Gamma_+\rightarrow3\sqrt{2}/4
\nonumber\\
\Gamma_{\rm rel}\rightarrow\Gamma_-/\sqrt{2};\qquad 
w_+/\rho_+\rightarrow2\sqrt{2}\Gamma_-/3
\eqe  
[using Eq.~(\ref{fixbland}]. For a parallel shock, where $b_-=b_+$,
the Alfv\'enic Mach number is
\eqb
M_{\rm A\pm}&=&{v_{\pm}\over b}\left({w_{\pm}\over\rho_-}+b^2\right)^{1/2}
\eqe
A switch-on shock occurs if the upstream flow is super-Alfv\'enic and
the hydrodynamic jump conditions give a sub-Alfv\'enic downstream speed, which
means, in the ultrarelativistic approximation,
\eqb
\sqrt{1+b^2}<\Gamma_-<\sqrt{3} b
\eqe
This parameter range is unusual, since it requires a plasma in which the magnetic
field energy density greatly exceeds the rest mass energy density, 
as well as a shock front whose normal is closely aligned with the 
upstream magnetic field.
\item
The ultra-relativistic approximation is more commonly encountered in connection
with perpendicular shocks, such as are thought to occur in the winds driven by
radio pulsars such as the Crab 
\cite{reesgunn74,kundtkrotscheck82,kennelcoroniti84,michel91}.
Here it is convenient to use the (Lorentz invariant) parameter
\eqb
\sigma&\equiv&{B^2\over4\pi w}
\nonumber\\
&=&{\rho_-b^2/w}
\eqe
which, in the case of a magnetic field orientated perpendicular to the flow, is
just the ratio of the energy flux carried by the electromagnetic field to that
carried by the particles.
Kennel \& Coroniti 
\cite{kennelcoroniti84} have given an approximate solution for strong,
ultrarelativistic perpendicular shocks:
\eqb
\Gamma_+^2&=&{8\sigma_-^2+26\sigma_-+17+\sqrt{64\sigma_-^2(\sigma_-+1)^2+20\sigma_-(\sigma_-
+1)+1}\over
16(\sigma_-+1)}
\nonumber\\
{b_+\over b_-}&=&{\rho_+\over\rho_-}\,=\,{\Gamma_-\over\sqrt{\Gamma_+^2-1}}
\nonumber\\
{w_+\over\rho_+}&=&
1+{\Gamma_-\over\Gamma_+}\left[1+\sigma_-\left(
1-{\Gamma_+\over\sqrt{\Gamma_+^2-1}}\right)\right]
\eqe
which goes over into the hydrodynamic solution (\ref{fixbland}) 
and (\ref{khformula}) for
$\sigma_-\rightarrow0$.
\end{enumerate}

\begin{figure}
\epsfxsize 15 cm
\epsffile{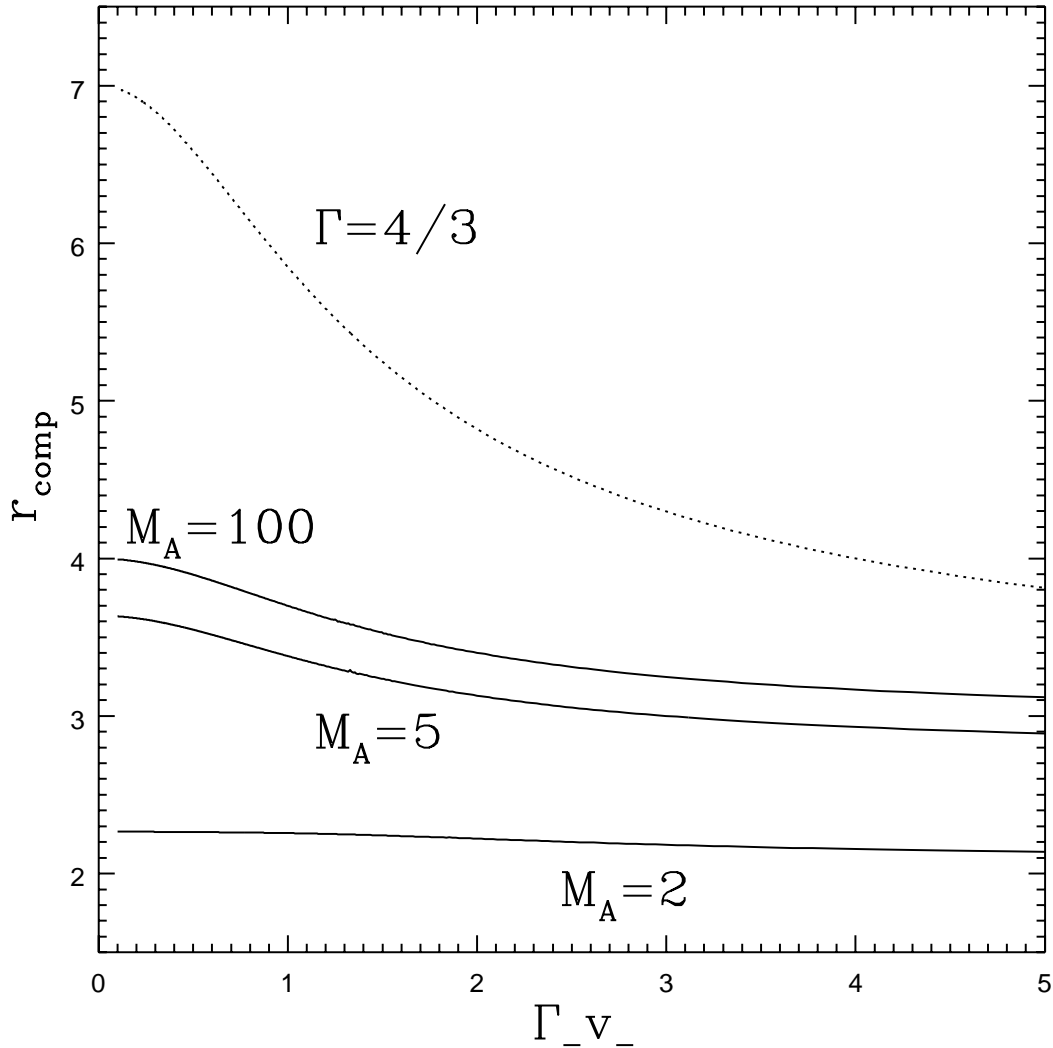}
\caption{
\protect\label{jump1}
The compression ratio of a strong oblique fast-mode 
shock front ($\Phi_-=45^{\rm o}$)
propagating into a
magnetised plasma. Various Alfv\'en Mach numbers $M_{\rm A}$ are shown,
together with the hydrodynamic approximation of
equation~(\protect\ref{fixbland}) (dotted line).
}
\end{figure} 
As pointed out by Kundt \& Krotscheck~\cite{kundtkrotscheck82}, the compression
which can be obtained in a relativistic shock decreases as the magnetic field
becomes more and more important dynamically. To illustrate this point we
present in figure~\ref{jump1} the jump conditions solved for the case of a magnetic field
at an angle of $45^{\rm o}$ to the shock normal in the upstream frame. The
shock is taken to be strong, in the sense that the pressure of the plasma
upstream is negligible. However, we allow for an appreciable magnetic pressure in
the upstream plasma, as measured by 
the Alfv\'enic Mach number, defined as $M_{\rm
A}=v_-/v_{\rm A}$, with the (relativistic) Alfv\'en 
speed given by equation~(\ref{alfvenspeed}). 
The Synge equation of state is used in these computations, and at Alfv\'en 
Mach numbers above 100, the result is close to the hydrodynamical case.
In addition, figure~\ref{jump1}
shows the hydrodynamic result for a strong shock using the fixed adiabatic
index approximation of equation~(\ref{fixbland}), and taking the
ultrarelativistic value $\hat\gamma=4/3$ (which, of course deviates
significantly from the Synge equation of state in the non-relativistic case). 

These computations were performed using an algorithm based on the description 
in \ref{mhdjump}. They extend the results presented by Ballard
\& Heavens~\cite{ballardheavens91} by using the Synge equation of state and
showing 
explicitly that the division
between super- and subluminal shocks used in that paper (and discussed in
Section~\ref{acceleration}) is unimportant for the jump conditions.
Similar results, but in terms of the parameter 
$\sigma_-=v_-^2/(M_{\rm A-}^2-v_-^2)$, 
are given for the ultra-relativistic perpendicular shock by
Kennel \& Coroniti \cite{kennelcoroniti84}. 
 

\section{Particle acceleration}
\label{acceleration}
Restricting ourselves to the picture of a shock front as a discontinuity in an
MHD flow, the crucial question to be answered before particle
acceleration can be discussed is the precise nature of the interaction between
accelerated particles, the plasma and its embedded magnetic field.
The standard starting point is orbit theory. Provided the embedded magnetic field varies slowly in time and space 
compared to the gyro-frequency and gyro-radius of the particle, the orbit can be described by the location of the 
gyro-centre (three spatial coordinates) the energy, the first adiabatic invariant or magnetic moment and the gyro-phase. The latter, which is the angle variable corresponding to the action variable \lq magnetic moment\rq, varies 
on the fastest time-scale. Correspondingly, it is assumed that the distribution function quickly adjusts itself to be 
independent of this coordinate. From this \lq unperturbed\rq\ orbit, the effect of small fluctuations in the 
electromagnetic field can be taken into account using quasi-linear theory. The result is a transport equation which 
is a second order partial differential equation describing diffusion in energy,
magnetic moment 
($=p(1-\mu^2)/B$, where $p$ is the momentum and $\mu$ the cosine of the 
\lq pitch angle\rq\ 
between the velocity vector and the magnetic field) 
and a coordinate perpendicular to the magnetic field. Explicit expressions for the diffusion 
coefficients, which are applicable in the rest frame of the plasma, 
have been computed by several authors 
\cite{hallsturrock67,melrose69,luhmann76,achatzetal91}. The 
most important wave mode responsible for the fluctuations is thought to be the Alfv\'en wave propagating parallel 
to the magnetic field, since it is only weakly damped 
\cite{lee82,lee83}. In this case the fastest 
diffusion time-scale is that associated with pitch angle scattering. Terms involving diffusion in energy and in 
position of the gyro-centre are smaller. In the next step, it is assumed that pitch angle diffusion is rapid enough to 
keep the particle distribution function close to isotropy, which enables the transport equation to be reduced to 
one containing only diffusion in the spatial coordinate along the field line, together with smaller terms describing 
cross-field and energy diffusion
\cite{jokipii66,hasselmannwibberenz70}. Neglecting diffusion in energy, one arrives at 
the well-known cosmic ray transport equation \cite{parker65,gleesonaxford67}. Applying it to a flow pattern 
which contains a non-relativistic shock front leads to the 
theory of diffusive shock acceleration 
\cite{krymsky77,axfordetal77,bell78,blandfordostriker78}.

This acceleration mechanism has been intensively studied 
(for reviews see Drury \cite{drury83} and Kirk et al.~\cite{kirketal94}). 
At a non-relativistic shock wave for
which $v_-$ and $v_+$ are each very much less than the particle speed 
$v_{\rm p}$, the particle distribution at the shock is, to lowest order
in $v_\pm/v_{\rm p}$, 
isotropic when viewed in either the upstream or downstream frames. A particle
which crosses from upstream to downstream and back again then undergoes a small
fractional increase in its momentum which is proportional to 
$v_{\rm rel}/v_{\rm p}$ and is given by $\Delta p/p=4(v_--v_+)/3v_{\rm p}$. 
The factor $4/3$ is a consequence of isotropy. The only
process which can stop this slow but inexorable acceleration is escape of the
particle from the vicinity of the shock. With $n_{\rm s}$ the number density
of particles at the shock the flux of (isotropic) particles of speed 
$v_{\rm p}$ crossing the shock is $n_{\rm s}v_{\rm p}/4$. Diffusion and 
isotropy guarantee that the number density of particles is constant in 
the downstream region so that the flux of escaping particles is then 
$n_{\rm s}v_+$. The probability of escape per shock crossing is then the 
ratio of the flux of escaping particles to those crossing the shock, 
$P_{\rm esc}=4v_+/v_{\rm p}$. The ratio of $P_{\rm esc}$ to $\Delta p/p$ 
determines the slope of the integral particle spectrum (i.e. the number of 
particles above a momentum $p$). The phase space density of particles 
then has a power law spectrum $f(p)\propto p^{-q}$ where 
$q=-P_{\rm esc}/(\Delta p/p)-3$. This gives a spectral index, {\it which 
depends on the compression ratio of the shock alone}, of $q=3r/(r-1)$ where
$r=v_-/v_+$. Note that this result does not depend on the level of scattering 
as long as that scattering is diffusive leading to an isotropic distribution 
when seen in each fluid frame.

Relativistic flow brings substantial changes to this picture, because 
the assumptions made in deriving the spatial diffusion equation are no 
longer valid. Thus, when $v_{\rm rel}$ is of order the velocity of 
light, the quantity $v_{\rm rel}/v_{\rm p}$ can no longer be small. The 
Lorentz boost linking the two frames significantly deforms the angular 
distribution of the particles, so that it cannot be nearly isotropic in 
both the upstream and downstream rest frames. Instead of assuming 
spatial diffusion, the angular dependence of the distribution function must be 
computed explicitly. 

Nevertheless, one can gain some understanding by exploiting a key property of the quasi-linear approach
 --  namely that diffusion across the field lines is ineffective -- despite the fact that a firm theoretical basis is 
lacking in the relativistic case. This property implies that, to a first approximation, a particle's guiding centre 
remains fixed to a given field line except at points where its orbit cuts the shock front. The kinematic 
implications are severe. The pattern of successive crossings and re-crossings of the shock front as a result of 
diffusion {\em along} the field lines, becomes progressively more difficult to maintain as the obliquity of the 
shock 
increases, since a diffusing particle has to chase not the shock front, but the point of intersection of the shock 
front with the field line to which it is tied. Re-crossing becomes impossible when the intersection point 
moves with a speed greater than that of light, 
i.e., $v_{\rm int}\equiv v_-/\cos\Phi_-\ge1$, and this divides shock fronts into two 
categories: subluminal ($v_{\rm int}<1$) and superluminal ($v_{\rm
int}>1$). The jump conditions themselves (see Section~\ref{shock}) do not
exhibit any special features at the change-over between sub- and superluminal shocks, however, there are 
interesting
differences when one considers the properties 
of the two types of shock under Lorentz transformations.
Thus, for subluminal shocks it is always possible to find a transformation to
the de~Hoffmann/Teller frame \cite{dehoffmannteller50} in which the electric
field vanishes everywhere. This is a convenient frame for computational
purposes. In the case of superluminal shocks, no such simple frame can be
found. The best that can be done is to transform to a frame in which the
magnetic field is perpendicular to the shock normal, the shock is stationary,
and the incoming flow is in the same plane as the shock normal and the
magnetic field \cite{drury83,begelmankirk90}. Generally, 
if one assumes the direction of the magnetic field in the upstream plasma 
has no causal connection with 
the shock front, so that it is randomly oriented with
respect to the shock normal, subluminal
shocks will greatly outnumber superluminal ones in subrelativistic flows. On
the other hand, in relativistic flows subluminal shocks will be rare, since
they require orientation of the magnetic field to within an angle of
$1/\Gamma_-$ of the shock normal.

At superluminal shocks, perturbations to the particle orbit are 
unimportant for acceleration when cross-field transport is neglected. 
However, even in the absence of fluctuations which scatter energetic 
particles, a shock front in a perfect fluid is capable of increasing the 
average energy of a distribution of particles incident upon it. Seen 
from the upstream rest frame, there exists an electric field in the 
downstream half-space simply because it is in motion (the 
${\bf- v\times B}$ field). A charged test 
particle can tap into this energy source if 
its gyrations through the regions of different magnetic field strength 
upstream and downstream cause it to drift along the shock front in the 
appropriate direction. This is usually called \lq shock-drift\rq\ 
acceleration. It can be analysed by tracing a large number of orbits 
numerically. The relativistic version of this mechanism has been 
investigated in \cite{begelmankirk90} and is described in 
Section~\ref{shockdrift}. 

In Section~\ref{analytic} we turn to the situation in which scattering 
does play an important role in the acceleration mechanism -- i.e., the 
relativistic generalisation of the diffusive shock acceleration 
mechanism -- and summarise the analytic and semi-analytic approaches to 
the problem. These involve the solution of the angular dependent 
transport equation assuming an explicit form for the pitch angle 
scattering coefficient. As such, they neglect cross field transport and 
apply principally to subluminal shocks. However, these methods may also 
be applicable to the case in which no a priori assumption is made 
concerning cross-field diffusion. If the direction of the embedded 
magnetic field is unimportant because a particle performs only a 
fraction of gyration between shock crossings (as in the case of an 
ultra-relativistic shock), or because the fluctuations in the field 
strength reach or exceed the amplitude of the embedded field, then 
transport by successive small angle deflections can be treated simply by 
replacing the pitch angle by the angle between the {\em shock 
normal} and the particle velocity \cite{gallantetal98}. 

Despite the advantages of a semi-analytic approach, 
Monte-Carlo simulation provides the most popular technique for investigating
acceleration at relativistic shocks. The formulation of the problem is 
conceptually simple, and the method permits one to look at a
variety of effects (such as synchrotron losses
and time dependence) which have so far defied an analytic approach. 
This work is discussed in Section~\ref{numerical}.

The ideal MHD approach, coupled with a simple transport prescription for
energetic particles 
allows us to make some predictions -- such as spectral indices
-- which are, at least in principle, accessible to observation. Currently, the
only work of which we are aware which goes beyond this picture concerns the
structure of relativistic perpendicular shocks, with particular application to
the Crab Nebula
\cite{hoshinoetal92,gallantetal92,gallantarons94}. The techniques 
involved -- one and 
two dimensional
particle-in-cell or hybrid simulations -- differ markedly from those employed
in the perfect fluid case, and are not discussed in this review.

\subsection{Shock-drift acceleration}
\label{shockdrift}

The dramatic increase in
surface brightness which can be produced by a relativistic shock front merely
as a result of the \lq compression\rq\ of the electrons
was pointed out by P.~Scheuer~\cite{scheuer89}. This raises the possibility
that observations of hot-spots in extra-galactic radio sources could be
understood even if the Fermi process with its emphasis on crossing and
recrossing of the shock front were unimportant.

To quantify this statement, consider a gas of relativistic electrons 
with an isotropic
distribution function in the local fluid frame. They achieve isotropy by
experiencing elastic scattering by slowly moving, low-frequency MHD waves,
which may be self-excited. The distribution in Lorentz factor $\gamma$ is
unaffected by the scatterings. 

The phase space density of particles contained in the
interval $\diff p$ around $p$ is conventionally assumed to be a power law:
\eqb
f(p)&=& C p^{-s}
\eqe
between a lower and an 
upper cut off: $p_{\rm min}<p<p_{\rm max}$.
The corresponding number density within this range of $p$ is
$n(p)=4\pi p^2 C p^{-s}$.
Let us assume that this distribution with $C=C_-$ 
accurately describes particles in a fluid
element which flows into an MHD shock. 

If the scattering events are so rapid that the length scale over which they
isotropise the electrons is much shorter than the length scale characterising
the thickness of the shock, then the
relativistic electrons react adiabatically. The Fermi process is unimportant in this case because scattering anchors 
the particles in the local fluid element and 
prevents them from repeatedly sampling the velocity difference between the 
upstream and downstream sides of the shock in between scatterings. 
The characteristics of
the transport equation are then
\eqb
p\rho^{-1/3}={\rm constant}
\eqe
where $\rho$ is the proper fluid density
(e.g., \cite{kirketal94}), so that 
downstream of the shock front one has
\eqb
f_+(p)&=&f_-[p(\rho_+/\rho_-)^{1/3}]
\nonumber\\
&=&C_-\left({\rho_+\over\rho_-}\right)^{s/3}p^{-s}
\eqe
Provided $p$ is far from the cut-off momenta, this can be written in terms of a
(proper) compression ratio for the electron distribution $R_{\rm e}$:
\eqb
R_{\rm e}&\equiv&{f_+(p)\over f_-(p)}
\label{electroncompr}\\
&=&
R^{s/3}
\label{compression}
\eqe
where $R$ is the proper compression ratio of the MHD fluid.

Even if they are not scattered whilst traversing the shock, charged 
particles may still be said to 
behave \lq adiabatically\rq\ in the sense that the first
adiabatic invariant of motion (the magnetic moment) 
is conserved. For this approximation to be valid, 
it suffices that the length
scale on which the magnetic field and fluid speed change (i.e., the shock
thickness) is long compared to the gyro-radius of the electron concerned.

Perhaps surprisingly, 
the magnetic moment is still conserved to a good approximation even if
the magnetic field changes discontinuously. This was pointed out by
Parker~\cite{parker58} and 
Schatzman~\cite{schatzman63} for perpendicular shocks and investigated
numerically by Decker~\cite{decker88}. From the work of Whipple et
al~\cite{whippleetal86} it is known that the magnetic moment is conserved
during encounter with a discontinuous change in the magnetic field at a 
perpendicular shock front provided the velocity of the front is small compared
to that of the electron. In the case of synchrotron emitting electrons, this
means $v_-\ll 1$. Away from the shock front (i.e., further than a gyro-radius from it) the particle distribution may 
relax slowly to isotropy as a result of pitch angle scattering. However, crossing and re-crossing is ruled out if the 
shock is superluminal, in which case this process operates undisturbed by the Fermi mechanism. The effective 
compression ratio for the electron distribution defined in equation (\ref{electroncompr}) is straightforward to 
calculate 
using Liouville's theorem applied to the drift-kinetic equation (e.g., \cite{begelmankirk90})
\eqb
R_{\rm e}&=&{1\over 2}R_{\rm B}^{(s-1)/2}{\pomega}^{-1/2}B_{\pomega}\left(
{1\over2},{s-1\over2}\right)
\\
&\approx&\sqrt{\pi\over2}R_{\rm
B}^{(s-1)/2}{\Gamma((s-1)/2)\over\Gamma(s/2)}
\nonumber\\&&{\rm for\ }R_{\rm B}>3
\label{vanderlaan}
\eqe
where $R_{\rm B}$ is the proper compression ratio of the magnetic field,
$\pomega=(R_{\rm B}-1)/R_{\rm B}$ and $B_{\pomega}(a,b)$ is an incomplete Beta function 
(\cite{abramowitzstegun72}, page~944).
The special case $s=4$ can be expressed in terms of elementary functions, and is given by van der Laan 
\cite{vanderlaan62}. If the magnetic field is dynamically unimportant, 
the magnetic compression $R_B$ is given by (\ref{khformula}).

At a relativistic shock, the magnetic moment of a charged particle is 
in general not even
approximately conserved, although for subluminal shocks there still exists a
small region of phase 
space for particles moving with the shock front within which the approximation
holds \cite{gieseleretal99}.
Notwithstanding this, significant progress can only be made by 
numerically following individual particle trajectories as they are carried over the shock front. Since those sections 
of the path between intersections are simply helices, the algorithm reduces to locating the intersection points and 
performing the relevant Lorentz transformations. Phase is now an important coordinate, so that results must be 
integrated over all initial phases, as well as initial pitch angles. The results presented in \cite{begelmankirk90} 
show 
that the electron compression ratio $R_{\rm e}$ is much larger at relativistic shocks than a naive extrapolation of 
the non-relativistic results (\ref{compression}) and (\ref{vanderlaan}) would indicate. In fact, for $\Gamma_{\rm 
rel}\gg1$, one has $R_{\rm e}\propto \Gamma_{\rm rel}^{s-2}$, compared to the asymptotic behaviours 
$R_{\rm e}\propto \Gamma_{\rm rel}^{s/3}$ and $R_{\rm e}\propto \Gamma_{\rm rel}^{(s-1)/2}$ 
predicted by equation (\ref{compression}) and equation (\ref{vanderlaan}), respectively. 
In connection with this result, it is interesting to note 
that at a strong relativistic shock front, the average Lorentz factor of {\em thermal} particles 
downstream is the same as their pre-shock value, as seen from the downstream rest frame. Essentially, the 
energisation in shock drift acceleration at an ultra relativistic shock front is only slightly greater than is achieved 
by isotropising the upstream particles using the downstream magnetic field and the static fluctuations associated 
with it.

Although each of the three mechanisms described in this section results in a strong enhancement of the energy 
density in accelerated particles, the boost in the energy of each individual particle is modest. The mechanisms 
work only when a population of particles with a non-thermal distribution flows in towards a shock front, and the 
characteristics, such as spectral index, of the incoming distribution are not changed. Thus, there is no 
characteristic spectral signature, nor is it possible by these mechanisms to accelerate a particle over several 
decades in energy. Nevertheless, these mechanisms would be the simplest ones to
incorporate into a numerical hydrodynamic or MHD simulation. 

\subsection{First order Fermi acceleration - analytic and semi-analytic methods}
\label{analytic}
In contrast to the shock-drift mechanism, the repeated crossings and re-crossings characteristic of first order 
Fermi acceleration enable large boosts in particle energy and, perhaps more importantly, produce a particle 
distribution with a characteristic spectral index.
Early on, it was thought that the increased escape probability in the
relativistic case would
mean that most of the accelerated particles would be swept away from the shock
never to return, which would severely reduce the acceleration \lq
efficiency\rq\ (i.e., cause the spectrum to fall off rapidly to higher
energies). However, this does not happen for two reasons. Firstly, the escape
probability is related to the speed of the shock in the downstream rest frame.
This quantity does not increase indefinitely as the shock becomes faster, but
remains lower than the relevant wave speed. In the case of a hydrodynamic
shock, the wave speed is that of sound, which cannot exceed $c/\sqrt{3}$.
Secondly, the acceleration per shock crossing also increases as the shock
becomes faster. In fact, there is no upper limit to this quantity, since a
particle which encounters a shock head-on (i.e., whilst moving along the shock
normal) and returns in the opposite direction suffers a fractional increase in
energy which is of the order of $\Gamma_{\rm rel}^2$. However, the average 
fractional energy gain is a sensitive function of the angular distribution 
of the particles, and, as we shall see, this adjusts itself such that the
competition between acceleration and escape results in a power-law spectrum
which does not differ radically from the non-relativistic value.

To find the spectrum of particles accelerated by the first order Fermi process
at a shock front it is necessary to solve the transport equation in the
upstream and downstream media and match the solutions at the shock front. In
contrast with the extensive analytic developments in the non-relativistic
theory, only a few special cases have been solved for relativistic flows. The 
simplest of these, which is outlined below is the determination of the characteristic power-law index of
test particles accelerated at a plane parallel shock front in which the fluid
flows along the shock normal. 
This was 
first attacked by
Peacock~\cite{peacock81}, who, however, did not solve the full transport
problem, but assumed a particular angular distribution
of particles in the downstream plasma. It was subsequently solved using an eigenvalue
expansion technique \cite{kirkschneider87a}.
This method has been generalised to treat
acceleration at oblique shocks, employing the additional assumptions of
conservation of magnetic moment upon crossing the shock, and neglect of
cross-field transport processes \cite{kirkheavens89}. 
Much harder spectral indices were found in these cases, largely as a result of
the accumulation of particles swept up into a precursor by 
repeated reflections from the shock front
\cite{ostrowski91,naitotakahara95,gieseleretal99}. These are ameliorated
somewhat when account is taken of the weakening of the compression ratio 
of an oblique shock front by magnetic pressure \cite{ballardheavens91}, 
as shown in figure~\ref{jump1}. 
Steeper spectra may also result if cross-field
transport becomes important.
 
\lq Modified\rq\
(parallel) shocks
-- i.e., those of finite thickness and a prescribed velocity profile formed the
subject of another generalisation of the method \cite{schneiderkirk89,kirkschneider89}. 
The technique described in the latter paper is
relevant for the problem of injection at non-relativistic shocks, where the
anisotropy of the particle distribution plays a key role \cite{malkovvoelk95}.
Adding a term describing large-angle scattering to the transport equation
allows further insight into the acceleration mechanism and its sensitivity to
the scattering operator using essentially the same technique
\cite{kirkschneider88}. Non-linear effects, which play a prominent role in
non-relativistic theory, have proved difficult to treat analytically -- the only
advance being the relativistic generalisation \cite{baringkirk91} 
of the classification of possible stationary solutions given in the diffusive
case by Drury and V\"olk~\cite{druryvoelk81}.   

Returning to the problem of test particles at a parallel shock,
in the presence of 
pitch angle diffusion described by the coefficient $D_{\mu\mu}$, 
but neglecting as usual diffusion in energy, the
equation to be solved is
\eqb
\Gamma_\pm(1+v_\pm \mu)
{\partial f\over\partial t} +
\Gamma_\pm(v_\pm+ \mu)
{\partial f\over\partial x}
&=&{\partial\over\partial\mu} D_{\mu\mu}{\partial 
f\over\partial\mu}
\label{reltrans}
\enspace,\eqe
where we have assumed the Lorentz factor of the particles is much larger than that of the flow and have 
accordingly replaced their velocity with the speed of light ($=1$).
The relatively simple form of (\ref{reltrans}) is a consequence of a mixed
coordinate system \cite{kirketal88} in which the cosine of the
pitch angle $\mu$ is measured in the local rest frame of the plasma, but the
space-time coordinates $x$ and $t$ refer to the rest frame of the shock front. 
Seeking a stationary solution in the shock frame, we separate 
the variables $\mu $ and $x$ to find an expression for
the general solution in terms of an 
eigenfunction expansion:
\eqb
f&=&\sum_{i=-\infty}^{\infty}
g_i(p)Q_i^{v}(\mu){\rm exp}(\Lambda^v_i x/\Gamma),
\label{eq4.1?}
\enspace,\eqe
where the $g_i(p)$ are arbitrary functions of momentum.
This solution is valid in both 
the upstream ($x<0$, $v=v_-$) or downstream ($x>0$, $v=v_+$) regions.
The eigenvalues $\Lambda^v_i$ and 
eigenfunctions $Q_i^{v}(\mu)$ appropriate to each half space
are solutions of the equation
\eqb
{\partial\over\partial \mu}
 D_{\mu\mu}{\partial\over\partial \mu}
Q_i^{v}(\mu)
&= &\Lambda^v_i(v+\mu) Q_i^{v}(\mu)
\label{eq4.2?}
\enspace.\eqe
together with the boundary conditions that the eigenfunctions be regular at the singular points $\mu=\pm1$
\cite{kirkschneider87a}.

The eigenvalue problem equation~(\ref{eq4.2?}) resembles the standard Sturm-Liouville problem familiar in 
many 
branches of physics 
and shares with it several convenient properties (\cite{ince56}, page 227), 
such as a discrete spectrum with eigenvalues ordered according to 
the number of 
nodes displayed by the corresponding eigenfunction. 
However, the special
feature of equation~(\ref{eq4.2?}) is that precisely at that value of $\mu$ 
at which particles are stationary with respect to the shock front, the \lq weighting function\rq\ 
$v_\pm +\mu$ changes sign. As a result, there are {\em two} families of eigenvalues and eigenfunctions for each 
value 
of $v$: one of them has positive, the other negative eigenvalues $\Lambda^v_i$. It is convenient to use the label 
$i$ to indicate this, choosing positive integers for the positive eigenvalues
and negative for the negative ones. The number of nodes possessed by the
eigenfunctions is then $|i|-1$. In addition to the two families of eigenvalues,
there exists a special isotropic eigenfunction with zero eigenvalue which is
labelled by $i=0$: $\Lambda_0^v=0$, $Q_0^v(\mu)=$constant. This just reflects
the property that pitch angle diffusion tries to drive any distribution towards
an isotropic one, which is the only stationary solution of the transport
equation in the absence of boundary conditions at finite $x$. The eigenfunctions are orthogonal and may be 
normalised such that
\eqb
\int_{-1}^{+1}{\rm d}\mu Q_i^v(v+\mu)Q_j^v&=&\eta_{ij}
\label{orthonormal}
\eqe
where $\eta_{ij}=0$ for $i\ne j$, $\eta_{ii}=1$ for $i\le0$ and $\eta_{ii}=-1$ for $i>0$.

In general, it is necessary to use a numerical technique to solve equation (\ref{eq4.2?}). Both a Galerkin method
\cite{kirkschneider87a,kirk88} and a shooting method \cite{heavensdrury88} have been employed. 
In the limiting cases of non-relativistic flow $v\ll 1$ and ultrarelativistic flow $v\rightarrow1$, approximate 
analytic expressions are available, which are described in \ref{appendixqj}

In the presence of a shock front at $x=0$, the general solution equation~(\ref{eq4.1?}) indicates that, in the 
upstream 
half-space ($x<0$), all terms containing eigenfunctions with $i<0$ diverge at large distance from the shock
($x\rightarrow-\infty$). Thus, for a physically admissible solution, the coefficients $g^-_i(p)$ must vanish for 
$i<0$. Furthermore, if we are interested in particles accelerated by the shock front, we can demand that the 
density vanish far upstream, which implies that the coefficient with $i=0$ must also vanish. Similar arguments 
can be applied to the downstream distribution, which, however, has non-vanishing density at $x\rightarrow\infty$.

\begin{figure}
\epsfxsize=15cm
\epsffile{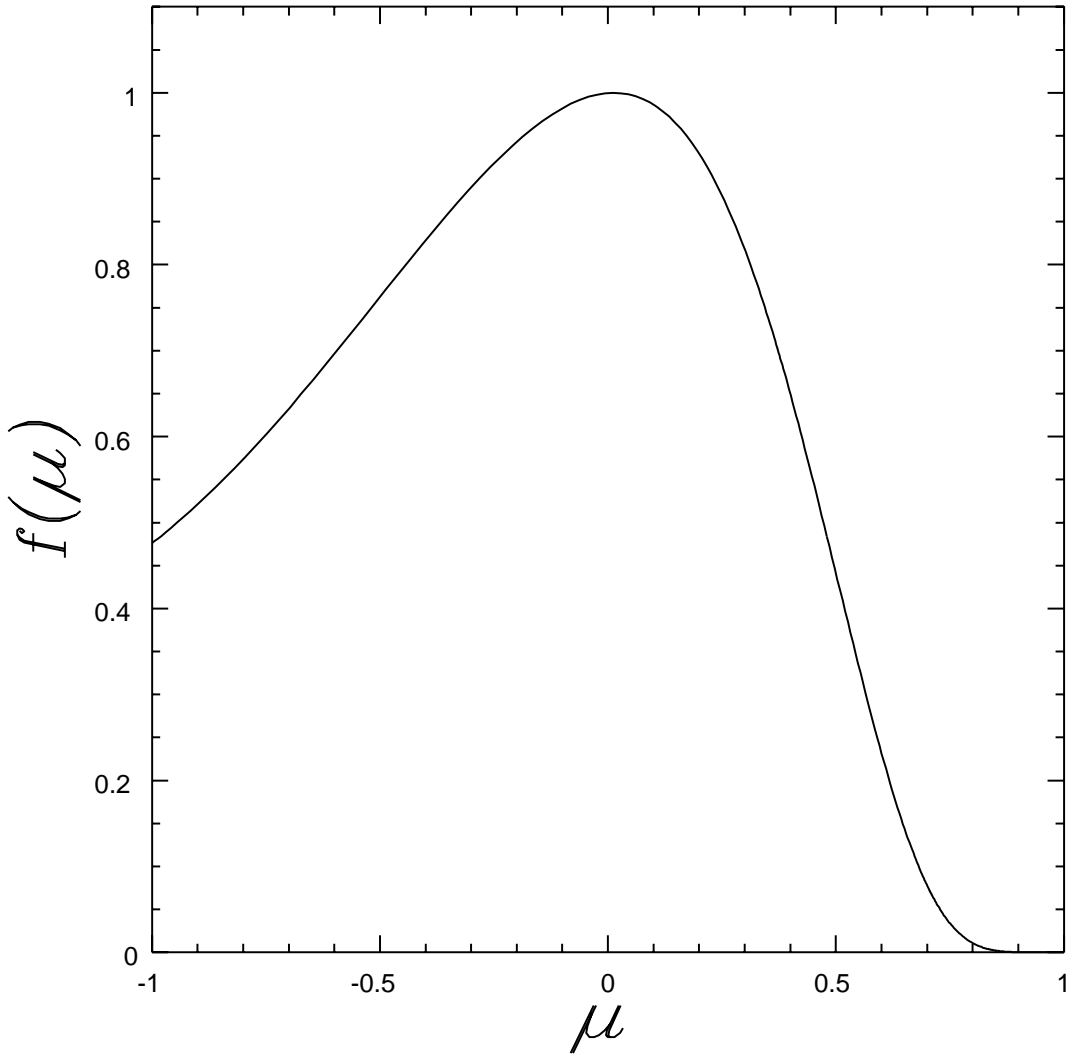}
\caption{
\protect\label{angulardist}
The pitch-angle distribution of accelerated particles at a parallel relativistic
shock front with $v_-=0.9$, and $v_+=0.37$, as a function of the cosine $\mu$ of the pitch angle
measured in the rest frame of the
downstream plasma. Isotropic pitch-angle diffusion is assumed and the normalisation of $f$ is arbitrary.
The depletion of particles with $\mu\approx1$, (those which 
move almost along the shock normal into 
the downstream plasma) arises because particles which move into the upstream plasma 
are overtaken again by the shock before undergoing substantial deflection
\protect\cite{gallantetal98}.}
\end{figure}
To find an approximation to the characteristic power-law index, one assumes $f\propto p^{-s}$, so that in 
equation 
(\ref{eq4.1?}) $g_i(p)=p^{-s}$, and uses as an Ansatz an approximation to the 
downstream distribution which fulfills the boundary 
conditions at $x\rightarrow\infty$.
\eqb
\tilde f(\tilde p,\tilde \mu,x)&=&
\tilde p^{-s}\sum_{i=-J}^0\tilde a_i
Q^{+}_i(\tilde\mu)
{\rm exp}(\Lambda^+_i x/\Gamma_+),
\label{eq4.62?}
\eqe
where the $\tilde a_i$ are constants, and we have truncated the expansion to
include only a finite number ($J+1$) eigenfunctions. 
We have also used the notation that the momentum and 
cosine of the pitch-angle measured in the downstream rest frame are $(\tilde p,\tilde\mu)$. They are related to 
those measured in the upstream frame $(p,\mu)$ by a Lorentz transformation
\eqb
\tilde p&=&\Gamma_{\rm rel} p(1+v_{\rm rel}\mu)
\label{eq4.59?}\\
\tilde\mu&=&{\mu+v_{\rm rel}\over1+v_{\rm rel}\mu},
\label{eq4.60?}
\eqe
The distribution function expressed as a function of downstream quantities is denoted by $\tilde f$.
On crossing the shock, particles are assumed not to undergo a sudden change in momentum (or direction of 
motion) so that Liouville's theorem demands continuity of the distribution. This condition reads
\eqb
f(p,\mu,x=0)&=&\tilde f(\tilde p,\tilde\mu,x=0)
\eqe
Accordingly,
the upstream distribution at the shock front, $x=0$, may be computed
from equation~(\ref{eq4.62?}) by substituting for $\tilde p$ and
$\tilde\mu$ using equations~(\ref{eq4.59?}) and (\ref{eq4.60?}).
One must now try to fit the upstream boundary conditions
as well as possible using the $J+1$ constants and the power-law
index $s$. Although $\tilde f$ of equation~(\ref{eq4.62?}) cannot be
required to be orthogonal to {\em all} the upstream eigenfunctions
which cause divergent behaviour at $x\rightarrow-\infty$, it can
at least be made orthogonal to those $J$ of them with the longest
range (i.e., smallest $|\Lambda|$). Projecting onto these gives:
\eqb
\lefteqn{\Gamma_{\rm rel}^{-s}\sum_{j=-J}^0
\int_{-1}^{+1}\,d\mu\,Q^{-}_i(\mu)(v_-+\mu)(1+v_{\rm rel}\mu)^{-s}
Q^{+}_j(\tilde\mu)\,\tilde a_j}\nonumber\\
&=&0
\label{eq4.63?}
\eqe
for $i={-J}\dots 0$.
The condition that the set of $J+1$ homogeneous 
equations~(\ref{eq4.63?}) possess a non-trivial solution 
is sufficient to determine
the unknown power-law index $s$. Once this is determined, the coefficients
$\tilde a_j$ and hence the angular dependence of  the distribution follow.

An example of the angular distribution at a relativistic shock, as seen from the 
rest frame of the downstream plasma, 
is shown in Fig. \ref{angulardist}. This figure was computed using an 
isotropic pitch-angle 
diffusion coefficient $D_{\mu\mu}\propto 1-\mu^2$. As well as confirming that the 
the distribution is strongly pitch-angle dependent, Fig. \ref{angulardist} shows 
that very few particles travel in the direction $\mu=1$, i.e.,  
along the shock normal into the downstream region. 
The reason is that a particle which crosses into the 
upstream plasma undergoes relatively little deflection before 
being caught again by the relativistically moving shock.

\begin{figure}
\epsfxsize=15 cm
\epsffile{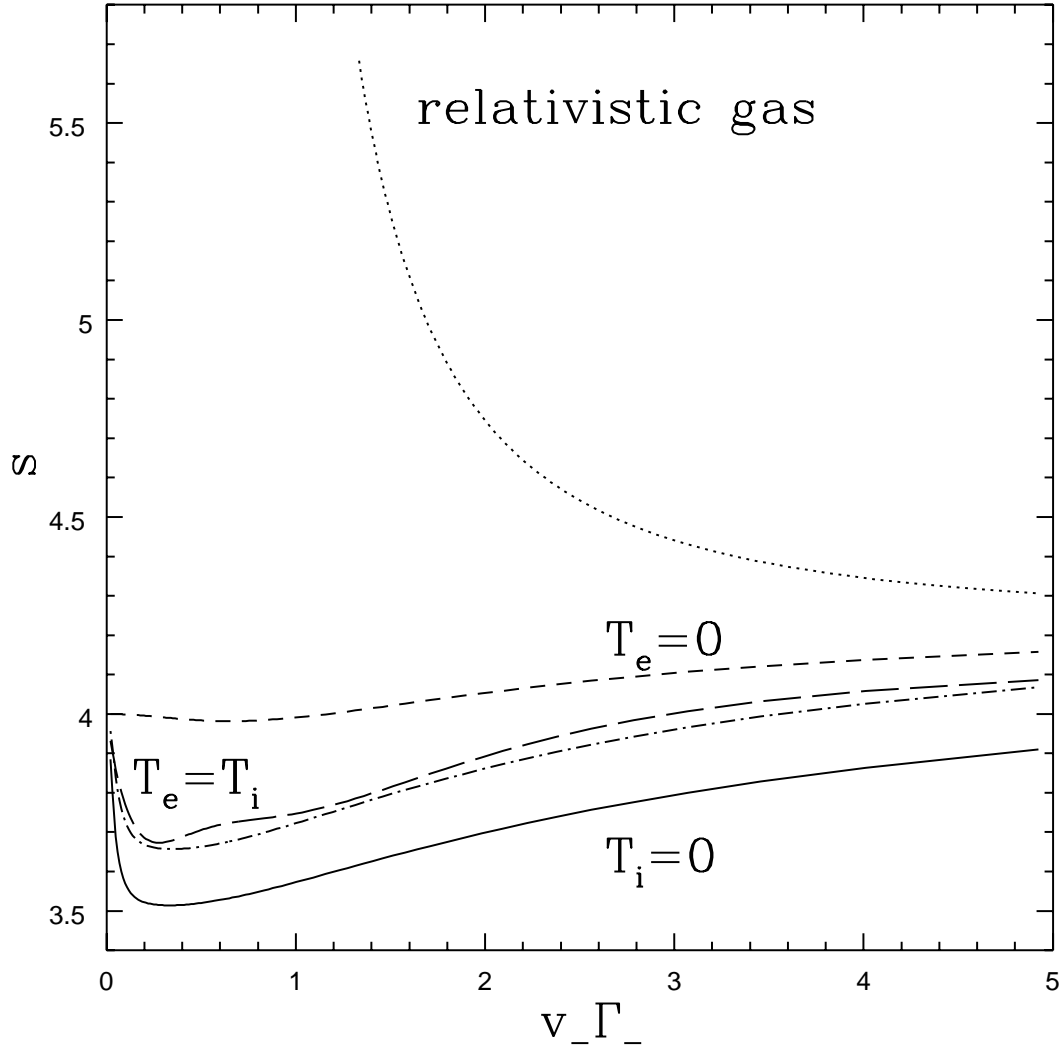}
\caption{
\protect\label{sindex}
The spectral index as determined using the $Q_J$ method for a 
parallel relativistic shock in plasma with various equations of state.
Plotted on the abscissa is the 
$x$ component of the four
velocity of the upstream plasma $v_-\Gamma_-$.
The dotted line corresponds to a relativistic gas with jump conditions
given by equation~(\protect\ref{simplejump}). The dashed line depicts a gas 
consisting of fully ionised hydrogen and helium (25\% by mass) with the
electrons assumed cold ($T_{\rm e}=0$); 
the solid line the same gas but with cold ions
($T_{\rm i}=0$, 
i.e., dominated by electron pressure). Two sets of results are shown for
complete thermodynamic equilibrium
($T_{\rm e}=T_{\rm i}$),  
those of Kirk \protect\cite{kirk88} (dashed-dotted line) for
the same composition as the other cases and those of 
Heavens \& Drury \protect\cite{heavensdrury88} for a proton/electron gas
(long dashed line), as given by equation~(\protect\ref{hdfit}).
}
\end{figure}

Results for the power-law index $s$ obtained by Kirk~\cite{kirk88} 
are summarised in Fig.~(\ref{sindex}),
where the characteristic spectral index is plotted as a function of 
$v_-\Gamma_-$ for
parallel shocks having various equations of state. The plasma composition in
this case is approximately primordial, with 25\% helium by mass.
Heavens and Drury~\cite{heavensdrury88} have given approximate formulae for
$s$ which they found by fitting their results in the range $0<v_-<0.98$. For
isotropic pitch angle diffusion and a gas consisting of electrons and protons
in thermodynamic equilibrium, they find for a strong shock
\eqb
s&\approx&3.99 - 3.16 v_- + 10.86 v_-^2 - 15 v_-^3 + 7.46 v_-^4
\label{hdfit}
\eqe
This fit is also shown in Fig.~\ref{sindex}. 
The agreement between the two treatments is clearly very
good; the minor differences probably being due to slight variations in the jump
conditions brought about by the different plasma compositions.

\begin{figure}
\epsfxsize=15 cm
\epsffile{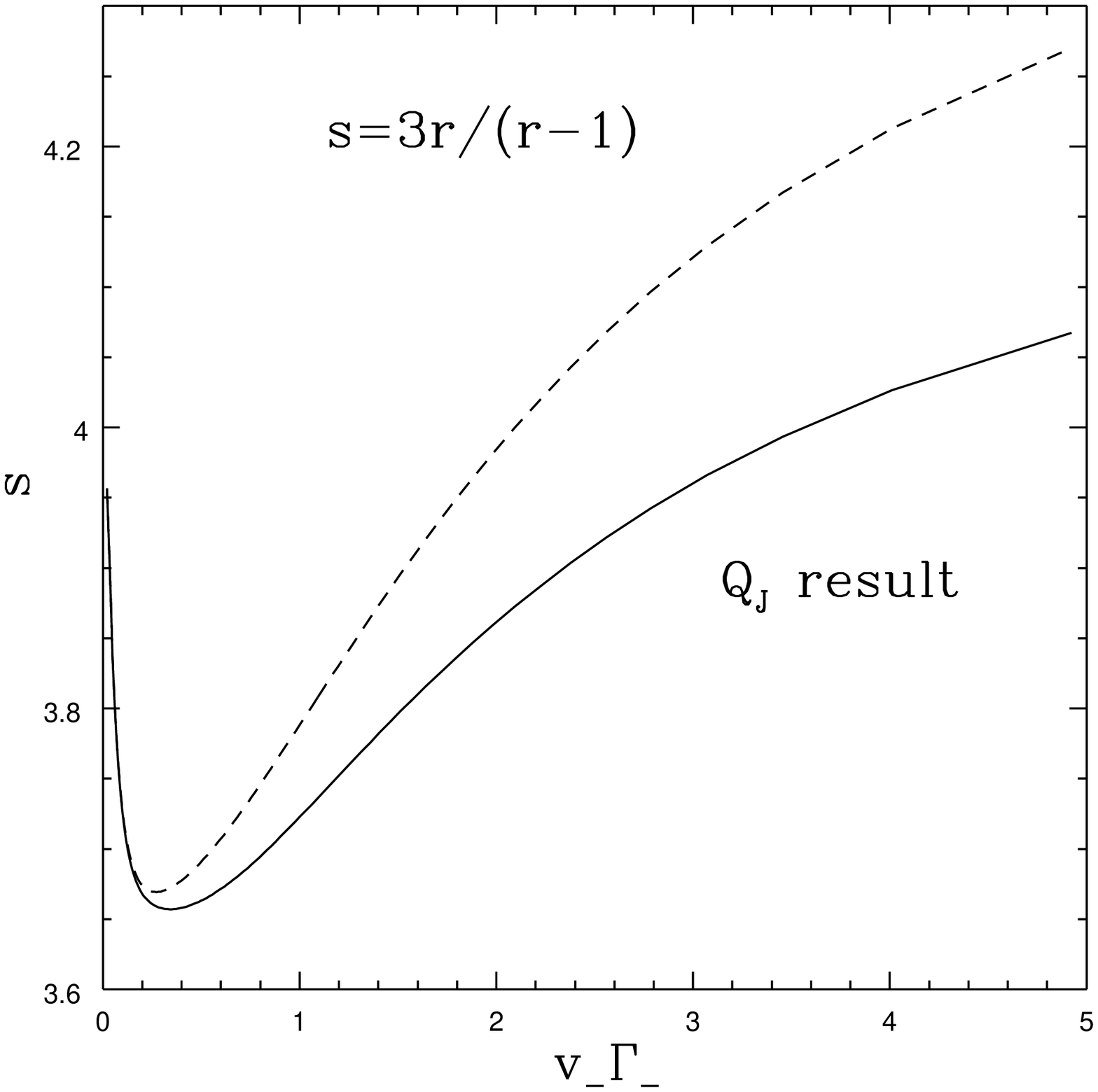}
\caption{
\protect\label{compare}
The spectral index as determined using the $Q_J$ method (solid line)
for a 
parallel relativistic shock in plasma (25\% helium) 
in full thermodynamic equilibrium.
The dashed line corresponds to the non-relativistic formula for the spectral
index: $s=3r/(r-1)$, where $r=v_-/v_+$ is the compression ratio}
\end{figure}

The main features of the
plot can be understood as simply being due to the different compression ratios
which arise when the shock fronts become relativistic. 
Nevertheless, significant deviations from a naive extrapolation of the non-relativistic
formula
$s=3r/(r-1)$ are found already at quite low speeds, as is shown in 
Fig.~(\ref{compare}). 

\begin{figure}
\epsfxsize=15 cm
\epsffile{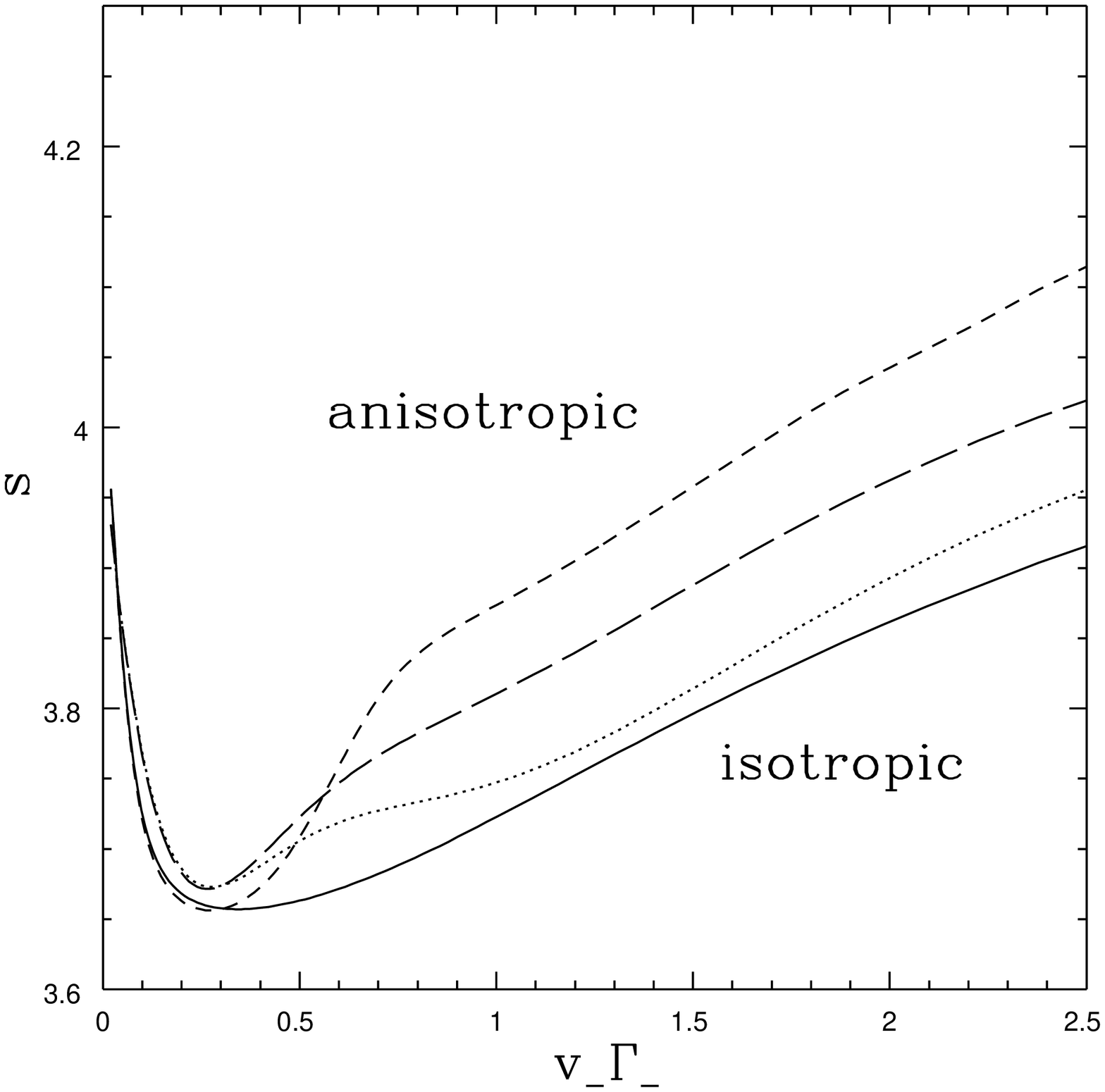}
\caption{
\protect\label{anisotrop}
The effect of anisotropic pitch angle diffusion on the spectral index produced
by a parallel relativistic shock front. The full line (isotropic) and the short
dashed line (anisotropic)
show the results of the $Q_J$ computation \protect\cite{kirk88}, with, for the
latter, pitch angle diffusion given by equation~(\protect\ref{padiffan}).  
The spectral index found by Heavens and 
Drury \protect\cite{heavensdrury88} is shown by the dotted line (isotropic)
and long dashed line (anisotropic). The pitch angle diffusion coefficient
in the latter case
is given by equation~(\protect\ref{padiffanhd}). Note that 
\protect\cite{kirk88} and \protect\cite{heavensdrury88} assume 
slightly different
compositions for the plasma.}
\end{figure}

In contrast to the diffusive case, the value of the spectral index $s$ 
for relativistic shocks depends 
on the functional form of the pitch-angle diffusion
coefficient $D_{\mu\mu}$. This point has been investigated by Heavens and Drury
\cite{heavensdrury88} and by Kirk \cite{kirk88}. For example, if 
pitch-angle scattering through the point $\mu=0$ is severely
restricted \cite{voelketal74}, the spectrum is steepened. 
Figure~\ref{anisotrop} illustrates this for pitch-angle scattering given by
\eqb
D_{\rm \mu\mu}&\propto&(1-\mu^2)\mu^{q}\qquad {\rm for\ }|\mu|>\epsilon
\nonumber\\
D_{\rm \mu\mu}&=&{\rm constant}\qquad{\rm for\ }|\mu|<\epsilon
\label{padiffan}
\eqe
with $\epsilon=1/30$ and the index $q$, which corresponds to the power-law
 spectrum of the turbulent wave-energy in the quasi-linear theory
equal, taken in this example to be 2.
Heavens and Drury, on the other hand adopt the prescription
\eqb
D_{\rm \mu\mu}&=&(1-\mu^2)(\mu^2+0.01)^{1/3}
\label{padiffanhd}
\eqe
which roughly corresponds to the quasi-linear result in the presence of
Kolmogorov turbulence.

To date, the method has been applied to shocks moving with a maximum Lorentz
factor $\Gamma_-=5$. Although the results in Fig.~\ref{sindex} seem to indicate
a convergence to a value around $4.2$ -- a result also found and commented upon
by Heavens and Drury~\cite{heavensdrury88} -- there is no analytic guarantee
that the asymptotic limit either exists or is approached
smoothly. Notwithstanding this, recent numerical results
(\cite{bednarzostrowski98,gallantetal98} -- see section~\ref{numerical}) 
also find convergence to $s=4.2$ for very large Lorentz factors.

\subsection{Numerical treatments of acceleration}
\label{numerical}
\subsubsection{Simulation of magnetic field}
The transport of charged particles in a relativistic flow can be simulated in a
number of different ways. At perhaps the most basic level, the entire
electromagnetic field can be specified in the relevant regions of space, and
the particle orbits integrated directly. The basic idea behind this approach is that the waves responsible for the 
transport of energetic particles are low-frequency, MHD modes, with phase velocities much smaller than the 
velocity of the shock front. In this case, it seems a reasonable approximation to neglect any time dependence of 
the electromagnetic field, apart from that associated with the fast-moving discontinuity which is the shock front 
itself. A single realisation of a stochastically generated static magnetic field configuration is then sufficient to 
determine the motion of all particles. 

This method has been employed by two groups to investigate particle acceleration at relativistic shocks 
\cite{ballardheavens92,ostrowski93}, with contradictory results.
There are several problems with the method. The most severe of these is the limited dynamic range 
available for simulation of the stochastic magnetic field. Ballard \& Heavens \cite{ballardheavens92} use a 
3-dimensional turbulent field with vanishing average component, generated by waves of random phase and 
intensity over a wavelength range of a factor of 100. Thus, a particle which during acceleration changes its 
gyro-radius by more than this factor will feel the effects of the limited dynamic range of the wave turbulence. Such a 
small energy range means that the characteristic index of the power-law spectrum expected of a realistic system 
can be 
determined only very approximately. Ostrowski \cite{ostrowski93}, on the other hand, uses a larger dynamic 
range, but employs only three monochromatic turbulent waves of stochastic amplitude but fixed phase within this 
range. He presents results for several orientations of the average magnetic field with respect to the shock normal.
However, it is not clear that the acceleration process is independent of the number of wave modes used.

Inherent in each of these methods is another difficulty: suppose that we 
succeeded in realising a stochastic 
magnetic field over a sufficiently wide range of length scales. How can we be  
sure that the numerical orbit 
integration reproduces the particle trajectories with sufficient accuracy? 
At first sight, this does not appear to be a 
serious problem. However, recent studies of anomalous 
transport regimes in fusion plasmas (e.g., Rax \& White 
\cite{raxwhite92}) and in the non-relativistic acceleration problem 
\cite{kirketal96} show that the property of 
reversibility of a particle trajectory in a static magnetic field is both 
difficult to achieve numerically and important 
for the determination of the characteristic spectral index in shock 
acceleration. The problem can be understood by 
considering the situation when long wavelength fluctuations create a local 
magnetic configuration in which a 
particle is \lq trapped\rq. When such a structure passes over the shock front,
the energy gained by a trapped 
particle depends strongly on how effectively the particle is confined 
\cite{achterberg88}. Physically, one might 
expect the perturbing effect of non-static field fluctuations to play a role.
This can be interpreted as an effective 
\lq decorrelation time\rq\ over which the particle forgets about a 
particular realisation of the static stochastic field.
Numerically, decorrelation is likely to be determined by the finite accuracy 
with which particle paths are computed. Because of 
the stringent requirements on accuracy, it has proved particularly difficult 
to devise numerical schemes which 
allow the study of the anomalous transport properties which arise on a 
time-scale shorter than the decorrelation time \cite{raxwhite92}.

\subsubsection{Simulation of particle transport}
An alternative, but perhaps less fundamental, approach to numerical simulation of acceleration is to make an 
assumption not about the stochastic field itself, but about the transport properties which it produces in the 
particles which are to be accelerated. This was a crucial step in
developing the original analytic diffusive shock acceleration models. It results in a
kinetic equation governing the single particle distribution function which
contains a collision operator. 

The basic idea of the Monte Carlo method 
is to find a way of constructing a stochastic trajectory whose
distribution obeys the desired transport equation. Then, by repeating the
procedure a large number of times, the distribution itself can be 
constructed approximately. 
This approach to solving transport problems has a long tradition, particularly
in the fields of neutron transport and radiative transport.
The transport of charged particles moving in a turbulent magnetic field
is, however, qualitatively different, since the turbulence acts continuously to
deflect the particle, whereas the scatterings of a neutron or photon are
impulsive events which act on the trajectory 
during short time intervals between which are long
periods of undisturbed flight.  
The differences are reflected in the mathematical properties of the 
collision operator contained in the transport equation. The pitch-angle
diffusion operator 
\eqb
{\cal C}[f(\mu)]&\equiv& 
{\partial\over\partial\mu} D_{\mu\mu}{\partial 
f\over\partial\mu}
\label{padiff}
\eqe
results in a transport equation which is
a second order differential equation, 
for example that given in equation~(\ref{reltrans}). 
This describes the continuous deflection of the particle by an infinite
succession of infinitesimally small changes in pitch angle.  
On the other hand,
the transport of photons and
neutrons is described by scattering through a finite angle. In the simplest
case of elastic scattering, one has, ignoring for the moment the gyro-phase,
\eqb
{\cal C}[f(\mu)]&\equiv&
\nu_{\rm scatt}\left[\int_{-1}^{1}\diff\mu' p(\mu,\mu')f(\mu')-f(\mu)\right]
\label{largeangle}
\eqe
where $\mu_{\rm scatt}$ is the scattering frequency and $p(\mu,\mu')\diff\mu$ 
is the probability that in a single scattering a particle with pitch angle cosine
$\mu'$ is scattered into the range $\diff\mu$ around $\mu$. 

The spectral
properties of these two operators are quite different \cite{kirkschneider88},
as is the approach to constructing a corresponding stochastic trajectory.
In the first case, one formally makes use of the Ito calculus (e.g., 
Gardiner~\cite{gardiner83}) to write a stochastic differential equation
(SDE). The appropriate 
stochastic trajectory is then found by numerically integrating the
SDE. An intuitive picture of this procedure is the following:
to proceed from one point on a
stochastic trajectory (labelled by, say $x_i,\mu_i$) to the next,
one must solve the transport equation under the initial condition 
$f(x,\mu,t=0)=\delta(x-x_i)\delta(\mu-\mu_i)$. For small changes in $x$,
this can be done approximately by assuming that $\mu$ changes only
slightly. Expanding equation~(\ref{reltrans}) in powers of $\mu-\mu_i$, 
one finds
\eqb
\Gamma_\pm(1+v_\pm\mu_i){\partial f\over\partial t}+
\Gamma_\pm(v_\pm+\mu_i){\partial f\over\partial x}&=&
D_{\mu\mu}(\mu_i){\partial^2\over\partial\mu^2}f+
D'_{\mu\mu}(\mu_i){\partial\over\partial\mu}f
\eqe
where $D'=\diff D/\diff\mu$. The substitutions 
$\Delta=(t-t_i)/[\Gamma_\pm(1+\mu_iv_\pm)]$, 
$\Xi=(x-x_i)/[\Gamma_\pm(v_\pm+\mu_i)]$
and $\eta=\mu-\mu_i+D'\Delta$ 
reduce this to the heat conduction equation, and 
the solution is easily seen to be
\eqb
f[x,\mu,t=t_i+\Gamma_\pm(1+\mu_iv_\pm)\Delta])&=&
{\delta(x-x_i-\Gamma_\pm(v_\pm+\mu_i)\Delta)
\over\sqrt{\pi D_{\mu\mu}\Delta }}
\nonumber\\
&&\exp\left[-(\mu-\mu_i-D'_{\mu\mu}\Delta )^2/(D_{\mu\mu}\Delta )\right]
\label{gauss}
\eqe
The next point on the trajectory is found by setting a sufficiently small
time step $\Delta$ and choosing a new stochastic 
value of $\mu$ from the
Gaussian distribution of equation~(\ref{gauss}). This method, together with
specialised techniques for enhancing the statistical significance of the
results (such as 
the \lq splitting\rq\ technique) was applied to the the
acceleration problem for particles undergoing synchrotron losses in
\cite{kirkschneider87b}. The Ito approach to more general problems including
second order Fermi acceleration has been presented by Kr\"ulls \& Achterberg
\cite{kruellsachterberg94}.

Whereas the Fokker-Planck operator stems from the quasi-linear theory of plasma
turbulence, as discussed in Sect.~\ref{analytic}, 
there is currently no theoretical justification for a large-angle
scattering operator such as equation~(\ref{largeangle})
to describe charged particles propagating in a turbulent
plasma. Nevertheless, motivated by the qualitative appearance of particle
trajectories in simulations of strong plasma turbulence 
\cite{zachary87}, which show sudden, large-angle
deflections, 
a mixture of the two operators has been investigated analytically
\cite{kirkschneider88}. The motivation to perform 
Monte-Carlo simulations using the large-angle
scattering operator is even stronger, since it is 
both relatively simple to implement and modest in its use of
computer time. 
A stochastic trajectory is constructed by first choosing a scattering
time distributed exponentially in the usual manner: 
$\Delta t=-\nu_{\rm scatt}{\rm ln}(\xi)$, where $\xi$ is a random number
distributed uniformly on the interval $[0,1]$. During this interval, 
the particle propagates with unchanged pitch angle from 
position $x_i$ to $x_{i+1}=x_i+(v_\pm+\mu_i)\Delta t$. 
The new value of $\mu=\mu_{i+1}$ is then found
from the distribution $p(\mu,\mu_i)$.
In the non-relativistic case, this 
method has proved highly popular (for a review, see~\cite{jonesellison91}).
Despite the lack of a theoretical background, the results it yields agree 
with solutions found using the diffusion approximation. This is because
the distribution function in the non-relativistic case 
is close to being isotropic. Independent of the mathematical structure of
the collision operator, it is then always possible to derive the equation of
spatial diffusion, so that a Monte-Carlo simulation 
using equation~(\ref{padiff}) must yield the same results as one performed 
using the more economical operator equation~(\ref{largeangle}). 

Once the motion becomes relativistic (or, more generally, once the fluid speed
approaches the particle speed), anisotropies
arise, and, with them, differences between the pitch-angle diffusion and
large-angle scattering cases \cite{kirkschneider88,ellisonetal90}. 
Nevertheless, large-angle
scattering with complete isotropisation at each scattering (i.e.,
$p(\mu,\mu')=0.5$)
has been used to estimate the acceleration time at relativistic shocks
\cite{quenbylieu89,ellisonetal90,lieuetal94,bednarzostrowski96},  
a quantity which is needed in order to estimate the
maximum energy cosmic rays which can be produced by the relativistic jet of a
radio galaxy. It is, of course, possible to choose the large angle scattering
operator such that not only the pitch angle is affected, but also the phase
and, more importantly, the position of the guiding centre. Particles can then
move across field lines as a result of scattering, mimicking the
quasi-linear process of cross-field diffusion. This effect has been studied in
connection with acceleration at non-relativistic oblique shocks
\cite{ellisonetal96}. 
If, in addition, the scattering is chosen to be strongly peaked in the forward
direction, then pitch-angle diffusion is simulated. In this way, 
a comparison of the effects
of pitch angle diffusion and large-angle scattering has
been made in investigations of relativistic, oblique shocks, both with 
\cite{naitotakahara95,gieseleretal99} and without 
\cite{ostrowski91} explicit
implementation of the assumption of magnetic moment conservation.

Monte-Carlo
simulations have also been performed for highly relativistic shocks by 
Bednarz \& Ostrowski \cite{bednarzostrowski98} using upstream Lorentz factors
up to roughly 240. They find the spectral index of accelerated particle converges to
the value $s=4.2$, independent of the orientation of the magnetic field, provided both 
pitch angle scattering and cross-field diffusion are permitted.
However, in the ultra-relativistic limit, a particle which manages to cross a
shock front from the downstream side into the upstream flow is very rapidly
overtaken again once it is deflected. In fact it, can perform
only a small fraction 
($\sim1/\Gamma_-$) of a gyration about the magnetic field line, unless the
direction of the field is exactly along the shock normal \cite{gallantetal98}.
In this case, the upstream transport cannot be represented by large-angle
scattering. Instead, a combination of motion in a uniform field, and diffusion
in angle due to fluctuations in the field on length scales much {\em shorter}
than a gyro-radius arises. If the field fluctuates rapidly, one would expect to
recover the operator equation~(\ref{padiff}), where now the quantity $\mu$ is
interpreted not as the cosine of the pitch angle, but as the cosine of the
angle between the particle velocity and the shock normal. 
Gallant et al \cite{gallantetal98} have extended the
method to the ultra relativistic limit ($\Gamma_-\rightarrow\infty$) and
considered both the case of diffusion in angle and scatter-free deflection by a
uniform field. The corresponding power laws are $s=4.25$ and $4.3$
respectively, in reasonable agreement with Bednarz and Ostrowski, despite
differences found in the angular distribution of the particles. This result
is particularly encouraging for those theories of gamma-ray burst sources 
which use a relativistic blast wave to accelerate the particles, since it is
close to the index of the particle spectrum required to produce afterglow 
spectra \cite{waxman97,galamaetal98}. 


\section{Discussion}
\label{discussion}

At the outset of this paper we pointed out that 
recent observations of blazars and gamma-ray bursts 
(GRBs) have each given renewed impetus to the study of particle 
acceleration in relativistic flows. We now address the issue of 
to what extent the results of these studies have been 
brought into contact with the observations
of such objects. 

Shock fronts arise when the 
relativistic flows encounter the ambient material. Variability in the central
\lq engines\rq\ of these sources may also result in internal shock waves, 
for example along the jets of AGNs. The shock jump conditions are then 
contained in the theory of section~\ref{shock}, at least at the MHD level, 
and, given some knowledge or model of the constituent matter and magnetic
field, the possible shock configurations follow. In principle it 
is then possible to apply the acceleration theory of section~\ref{acceleration}
to model the energisation of highly relativistic particles. 
However, although reliable computations of flow patterns are now available, the
most common approach to acceleration is still to  
assume a power law distribution of particles. 
In those instances where the 
acceleration is modelled at a deeper level it is usually with a simple
phenomenological equation 
for the distribution 
\cite{marcowithetal95,romanovalovelace97,levinson98,kirketal98}.

Before discussing the issues relating to jets and GRBs it 
should be noted that there exists in Nature one relativistic shock front 
which we are particularly well-placed to observe ions \cite{reesgunn74}. 
It is located 
in the Crab Nebula, and is responsible for thermalising the energy 
which streams from the central pulsar in the form of a relativistic flow 
of positrons and electrons, possibly contaminated by. 
It is thought that the shock is stationary and the magnetic field oriented 
perpendicular to the shock normal \cite{kennelcoroniti84}. 
Using $1{1\over2}\,$D numerical simulations, a determined effort has 
been made to understand both the plasma physics of the shock front 
and the connection with particle acceleration. 
However, 
this source seems fundamentally different to gamma-ray blazars and gamma-ray
bursts. Not only is it much less powerful, but it also 
does not display the rapid variability characteristic of the 
extra-galactic sources. In view of this, and because of the very different
technical approach employed, we do not attempt to synthesise this theory into 
our discussion, but refer the interested reader to the original papers 
\cite{hoshinoetal92,gallantetal92,gallantarons94}.

\subsection{Relativistic Jets}

The problem of how particles are accelerated to non-thermal energies in
relativistic jets has been discussed for a number of years 
\cite{begelmanblandfordrees84}. Diffusive acceleration at non-relativistic 
shocks is a possibility which appears to provide a reasonable picture of
acceleration at several jet hot-spots which emit synchrotron radiation.
In particular, the observed
softening of the spectrum towards higher frequencies has been 
compared with analytic 
computations which include energy losses by synchrotron radiation
\cite{meisenheimerheavens86,heavensmeisenheimer87,meisenheimerroser89}. 
However, both apparent superluminal motion 
and the rapid 
variability of emission from blazar jets imply substantial Doppler boosting,
which indicates that the non-relativistic theory is inadequate. 

The first attempt to model the radio emission of relativistic jets
was made by Wilson and Scheuer \cite{wilsonscheuer83}, who
performed numerical hydrodynamic simulations in which it was 
assumed that a fixed fraction of the energy thermalised at a shock front is
converted into relativistic particles and magnetic fields. This 
is a rather sweeping assumption, covering not only particle acceleration, but
also a dynamo process to convert part of the available energy into magnetic
flux. Although such processes may well occur, there is currently no theoretical
prospect of checking whether or not the assumed values of the 
parameters are realistic. Subsequent investigations have
chosen to inject particles at a particular point and simply allow them to be
compressed adiabatically at shock fronts according to Eq.~(\ref{compression})
\cite{gomezetal97,mioduszewskietal97,komissarovfalle97}. Since all of these
are hydrodynamical schemes, which do not follow the evolution of the 
magnetic field, equipartition was assumed. If correct at injection,
this assumption continues to hold through homologous
compressions or expansions, but it is violated at a shock front. 
Of course, if the field is in some sense random, the average effect of a shock may be 
similar to homologous compression, but this is not going to be the case for sources with 
a high degree of polarisation. Only a rapidly acting dynamo (or, in some cases,
reconnection) could then restore equipartition and thus rescue the
simulations. 
Dropping the assumption of
equipartition by assuming the fields to be frozen into the plasma (as in ideal
MHD), 
Matthews and Scheuer \cite{matthewsscheuer90a,matthewsscheuer90b} have computed the
emission patterns, once again assuming that relativistic electrons
are compressed adiabatically at shock fronts. Although they enable
interesting simulated
\lq radio maps\rq\ to be constructed, the 
major shortcoming of these simulations is that they do not include enough
of the physics of particle acceleration to be able to predict
spectra. 

The best developed model of blazar spectra is based on the picture of a shock
front moving down a jet \cite{marschergear85}. Both the magnetic field and 
the particle distribution vary with position and the observed radiation 
is a superposition from different parts of the jet 
\cite{hughesalleraller91,marschertravis96}. These models are generally 
called \lq inhomogeneous\rq. However, recent results concerning the rapid 
variability of blazar sources at all frequencies \cite{wagnerwitzel95} and 
the detection of their emission at energies up to at least $10\,$TeV
\cite{aharonianetal99} have provided new restrictions on the possible 
acceleration mechanisms. The observed simultaneous variations 
in X-rays and TeV gamma-rays indicate that a single population of particles in 
a relatively localised region is responsible. Consequently \lq homogeneous\rq\ 
models have been widely discussed \cite{dermerschlickeiser93,
ghisellinimaraschidondi96,steckerdejagersalamon96,mastichiadiskirk97,kirketal98} and
there is rough agreement on the parameters of the emission region.
Furthermore, simultaneous variability of emission at widely differing 
wavelengths is more easily interpreted in terms of directly accelerated 
electrons than in hadronic models \cite{mannheimwesterhoff96}, where the high
energy emission results from the acceleration of protons. 

In the case of blazars the synchrotron emission in the radio to infra-red range
typically displays a very hard power-law ($\alpha<0.5$). For homogeneous models
this means a flatter particle spectrum is needed 
than the canonical $s=4$ result of non-relativistic theory, since the cooling
time of the responsible particles is long, and the indices are related by
 $\alpha=(s-3)/2$. For Markarian~421, for example, a value of $s=3.7$ is required 
to give the correct index of $\alpha\approx 0.35$ for the synchrotron
emission. From Fig.~\ref{sindex}, we see that, in the case of parallel shocks, 
such values are obtained only for mildly relativistic speeds in plasmas in
which the electron pressure dominates. A perhaps more plausible alternative 
is to 
appeal to the fact that oblique relativistic shocks, through a combination of 
shock-drift and first-order Fermi acceleration, give much harder spectra 
ranging up to $s=3$ \cite{kirkheavens89}. This is due to the increased 
importance of reflections back into the upstream region from the shock front 
with increasing obliquity and shock speed, as discussed in section~\ref{analytic}. 

\subsection{Fireball models of GRBs}

The most popular, generic model for GRB sources is that of the fireball 
\cite{cavallorees78, goodman86, paczynski86} which involves the sudden 
release of energy in the form of an optically thick $e^{\pm}$ plasma. While
initially confined to a small volume, the plasma, and along with it the 
radiation, expands out to the point where the system becomes optically 
thin, releasing the electromagnetic radiation. The presence of a baryonic 
component can, however, result in the conversion of the initial internal 
energy into the kinetic energy of a thin shell of matter which expands 
relativistically, if such baryonic \lq contamination\rq\ is small, into the 
surrounding medium \cite{reesmeszaros92}. When this shell interacts with the 
ambient medium a relativistic shock front forms which ultimately dissipates the
kinetic energy of the ejecta into post-shock thermal energy and the production
of energetic particles. In qualitative terms at least this picture is 
analogous to that of an expanding supernova remnant where the theory of 
diffusive shock acceleration can be applied. However, although the hydrodynamics
of an expanding fireball have been well studied 
\cite{blandfordmckee76,blandfordmckee77,meszaroslagunarees93,piransheminarayan93, saripiran95}, 
the theory 
of particle acceleration at relativistic shocks, with the exception of some
Monte-Carlo studies, has not been applied to these sources. This is
also true of the case where a fluctuating Lorentz factor for 
the ejecta leads to the formation of internal shocks 
\cite{reesmeszaros94,saripiran97}. Likewise, arguments favouring a common origin for 
ultra-high-energy cosmic rays and GRBs \cite{vietri95, waxman95} have used
simple estimates of the acceleration processes which may take place.
If the problem of 
acceleration is essentially treated as one of {\it injection} where a value 
of $s=4$ is assumed and the total energy in energetic particles is 
parameterised as some fraction of the total fireball energy, estimates of the 
synchrotron and synchrotron self-Compton fluxes can be made 
\cite{meszarosrees93, dermer98} and compared with observations.
However, there are many important physical questions still unanswered. 
How does the environment and its equation of
state influence the spectrum?
Can a description
of acceleration at multiple, internal shock waves account for the variability
in GRBs? 

It is not hard to see why the acceleration part of the problem has been 
largely simplified in the literature for GRBs. Until recently, published
results were restricted to Lorentz factors of less than about~10,
whereas GRBs require values ten times larger. 
First results have now appeared
for the first-order Fermi process in 
the ultra-relativistic case \cite{bednarzostrowski98,gallantetal98} and  
report a characteristic value of the power-law index 
encouragingly close to that which the observations seem to 
indicate \cite{waxman97,galamaetal98,wijersreesmeszaros97}. As in the non-relativistic case, the
compression ratio at an ultra-relativistic shock is independent of the
shock speed. If, as the simulations suggest, the power-law index of
accelerated particles turns out also to be 
independent of the shock speed, modelling of individual events will 
become simpler, and 
the theory of particle acceleration 
may play an important role in advancing our understanding these sources. 


\ack
We are especially indebted to Y.~Gallant and A.~Heavens
for a careful reading of the 
manuscript and a number of important suggestions. 
We also thank A. Achterberg and J. Lyubarskii for 
stimulating discussions. This collaboration was supported by the 
TMR programme of the European Commission under the network
contract number ERBFMRX-CT-0168.

\appendix


\section{Hydrodynamic jump conditions}
\label{hydrojump}
Consider a fluid made up of a number of constituents
(electrons, protons and atoms of H and He, for example), labelled by the suffix
$i$, and assign to each a temperature $T_i$. 
For an ideal gas of 
monatomic, non-degenerate
constituents, the 
contribution $w_i$ of each of them to the total enthalpy density
is
\eqb
w_i&=&m_i n_i G(m_i/T_i)
\eqe
where $m_i$ is the rest-mass of the $i$'th constituent and $n_i$ its proper
number density. Note that we set the speed of light and Boltzmann's constant
equal to unity $k_{\rm B}=c=1$. The function $G(z)$ is given in terms of
modified Bessel functions of the first kind (McDonald functions):
\eqb
G(z)&=&{K_3(z)\over K_2(z)}
\eqe
(\cite{synge57}), and has the asymptotic expansion for low temperatures
\eqb
G(m/T)\rightarrow 1+{5T\over 2m}\quad{\rm as\ }T\rightarrow0
\eqe
and for high temperatures
\eqb
G(m/T)\rightarrow {4T\over m} + 1\quad{\rm as\ }T\rightarrow\infty
\eqe
Each constituent satisfies the ideal gas law
\eqb
p_i=n_i T_i
\label{idealgas}
\eqe
and the overall pressure, enthalpy density and energy density are simply
\eqb
p&=&\sum p_i
\nonumber\\
w&=&\sum w_i
\nonumber\\
e&=&w-p
\eqe
We further define the rest-mass density $\rho$ 
\eqb
\rho&=&\sum m_i n_i
\eqe
and the abundance (by mass) of the constituent $i$:
\eqb
\eta_i&=&m_i n_i/\rho
\eqe
so that $\sum \eta_i=1$.

The equation of state for this dissipation free (adiabatic) case
is then
\eqb
w&=&\rho\left[\sum \eta_i G(m_i/T_i)\right]
\label{eqofstate}
\eqe
together with a relationship between the $T_i$. (For example, in thermodynamic
equilibrium: $T_i=T$, or, for cold electrons/hot ions: 
$T_{\rm e}=0$, $T_{\rm H}=T_{\rm He}=T$.)

The speed of sound is simply given by $v_{\rm s}^2= \left(\partial p/\partial e\right)_s$
where $s$ is the specific entropy defined by the first law of thermodynamics:
$T\diff s=\diff(e/\rho)+p\,\diff(1/\rho)$. Using Eq.~(\ref{fixedgamma}),
and assuming $\adindex$ to be constant, we
obtain 
\eqb
v_{\rm s}^2&=&{{\adindex}p\over\rho}\left[{({\adindex}-1)\rho\over
({\adindex}-1)\rho+{\adindex}p}\right]
\label{soundvel}
\eqe
which yields the familiar non-relativistic result for $p\ll \rho$ and
gives $v_{\rm s}\rightarrow 1/\sqrt{3}$ for a relativistic gas.

The jump conditions Eqs.~(\ref{hydrojumpcond1}) and (\ref{hydrojumpcond2})
can be simplified by the introduction of the following notation:
let the quantities $\phi_\pm$ be defined by
\eqb
\Gamma_\pm&=&{\rm cosh}\phi_\pm\,.
\label{3.19?}\\
\noalign{\hbox{Then one has}}
v_\pm&=&{\rm tanh}\phi_\pm
\label{3.21?}
\eqe
The possibility of particle creation at the shock front may be taken into account by reserving the notation
$\rho_+$ for the density of conserved particles, and introducing the parameter $\eta$ 
to relate the total downstream rest-mass density $\rho_{{\rm total}+}$ to 
$\rho_+$:
\eqb
\eta&=&\rho_+/\rho_{{\rm total}+}
\label{etadef}
\eqe
If particle number is conserved, $\eta=1$, for small $\eta$, a large number of particles 
of non-zero rest-mass are created 
at the shock.
In terms of the energy per unit rest mass $\bar e_-=e_-/\rho_-$ and
$\bar e_+=e_+/\rho_{{\rm total}+}$
and the similarly
defined pressure per unit rest mass $\bar P_\pm$, one obtains
\eqb
\rho_+{\rm sinh}\phi_+ &=& \rho_-{\rm sinh}\phi_-
\label{3.22?}\\
(\bar e_+ + \bar P_+){\rm sinh}^2\phi_+ + \bar P_+ &=&
\eta{\rho_-\over\rho_+}\left[ (\bar e_- + \bar P_-){\rm sinh}^2\phi_-
+\bar P_-\right]
\label{3.23?}\\
(\bar e_+ + \bar P_+){\rm sinh}\phi_+\,{\rm cosh}\phi_+&=&
\eta{\rho_-\over\rho_+}{\rm sinh}\phi_-\,{\rm cosh}\phi_-
(\bar e_- + \bar P_-).
\label{3.24?}
\eqe
Elimination of $\rho_+/\rho_-$ leads, after a little manipulation, to
\eqb
\bar w_+{\rm cosh}\phi_+&=&
\eta\bar w_-{\rm cosh}\phi_-
\label{3.25?}\\
\noalign{\hbox{and}}
(\bar w_+{\rm cosh}^2\phi_+ - \bar e_+)&=&
\eta\bar e_-{\rm sinh}\phi_+\,{\rm sinh}\phi_-\left(1+{\bar P_-{\rm
coth}^2\phi_-\over\bar e_-}\right),
\label{3.26?}
\eqe
where $\bar w_\pm=\bar e_\pm + \bar P_\pm$.
The last term on the right-hand side of equation~(\ref{3.26?}) can be
neglected if one restricts one's consideration to strong shocks, i.e. to
shocks in which the upstream pressure is negligible. This
approximation may be extended by assuming the upstream medium is cold,
in which case the gas possesses only that energy attributable to its
rest mass: $\bar e_-=1$. Then, eliminating ${\rm cosh}\phi_+$, one has
\eqb
{\rm cosh}^2\phi_-&=&
{\bar w_+^2(\bar e_+^2-\eta^2)\over \eta^2(\bar e_+^2-\eta^2-\bar
P_+^2)}\,,
\label{3.27?}
\eqe
which is a generalisation to the case $\eta\ne1$ of a relation found by Peacock
\cite{peacock81}.
Alternatively, one may obtain from Eqs.~(\ref{3.25?}) and (\ref{3.26?})
the relation
\eqb
\bar e_+&=&\eta{\rm cosh}\phi_-{\rm cosh}\phi_+ - \eta{\rm sinh}\phi_-
{\rm sinh}\phi_+
\nonumber\\
&=&\eta{\rm cosh}(\phi_- - \phi_+)
\label{consavlor}
\eqe
The right-hand side of Eq.~(\ref{consavlor}) is just $\eta$ times
the Lorentz factor
$\Gamma_{\rm rel}$
associated with the relative velocity of the upstream 
fluid with respect to the
downstream fluid. 
Thus, in the absence of 
particle creation ($\eta=1$), the average energy per particle 
as seen from the frame of the downstream fluid 
is constant across a strong shock (in which $\bar e_-=1$). 
\cite{johnsonmckee71}.

For a given upstream velocity $v_-$,
equation~(\ref{3.27?}) must be solved
numerically to give the downstream parameters. The procedure is as
follows: $v_-$ determines ${\rm cosh}\phi_-$ through 
Eq.~(\ref{3.19?}). The right-hand side of (\ref{3.27?})
is a function of the
single parameter $\bar e_+$ (the average Lorentz
factor), since the pressure
$\bar P_+$ is given through the equation of state Eq.~(\ref{eqofstate}).
A root finding algorithm can therefore be
used to find $\bar e_+$. By means of equation~(\ref{3.25?}), which for strong shocks
specialises to
\eqb
\bar w_+{\rm cosh}\phi_+=\eta{\rm cosh}\phi_-,
\label{3.28?}
\eqe
one finds the downstream velocity $v_+$.

The ultra-relativistic case is best considered without imposing
the restriction to strong shocks.
The equation of state $p=e/3$ can then be
applied in both the
upstream and downstream regions.
Returning to equations~(\ref{3.25?}) and (\ref{3.26?}), one finds,
after a straightforward calculation,
\eqb
v_-v_+={1\over3}.
\label{3.32?}
\eqe

A useful analytic expression 
for strong shocks can be found if the equation of state
(\ref{fixedgamma}) is used \cite{blandfordmckee76}. The procedure is to
write the specific enthalpy in terms of $\Gamma_{\rm rel}$
using Eq.~(\ref{consavlor}):
\eqb
\bar w_+&=&{\adindex}(\eta\Gamma_{\rm rel}-1)+1
\eqe
Remembering that $\bar w_-=1$ for a strong shock, the ${\rm cosh}\phi_+$ term
of Eq.~(\ref{3.25?}) can be
rewritten as a function of $\phi_+-\phi_-$ and $\phi_-$ 
to give an expression for ${\rm tanh}\phi_-$. Writing this in 
terms of $\Gamma_{\rm
rel}={\rm cosh}(\phi_+-\phi_-)$ one finds
\eqb
\Gamma_-^2&=&
{{\bar w}_+^2(\Gamma_{\rm rel}^2-1)\over
{\adindex}(2-{\adindex})(\eta\Gamma_{\rm rel}-1)^2+2(\eta\Gamma_{\rm rel}-1)
+1-\eta^2}
\label{bland}
\eqe
Thus, given $\eta$ and 
the relative speeds of the upstream and downstream fluids, the
shock speed and downstream pressure can be written down.


\section{MHD jump conditions}
\label{mhdjump}

In solving the MHD shock jump conditions it is useful to define the components of $b_\pm^\mu$ and 
$u_\pm^\mu$ along the shock normal, $d_\pm\equiv b^\mu_\pm l_\mu$ and 
$a_\pm\equiv u^\mu_\pm l_\mu$. These Lorentz scalars take the values
$d_\pm=\Gamma_\pm(B_\pm/\sqrt{4\pi\rho_-})\cos\Phi_\pm$ and $a_\pm=\Gamma_\pm v_\pm$. With 
$n_\pm=\rho_\pm/\rho_-$ the jump conditions become
\eqb
\left[na\right]&=&0
\nonumber\\
\left[V^\alpha\right]&\equiv&\left[du^\alpha-ab^\alpha\right]=0
\nonumber\\
\left[W^\alpha\right]&\equiv&\left[anGu^\alpha+{n\over z}l^\alpha+
|b|^2\left(au^\alpha+{1\over 2}l^\alpha\right)-db^\alpha\right]=0
\eqe
where $z=\rho/p$.
These are a set of non-linear equations in the unknown downstream quantities. However, 
they can be reduced to one transcendental equation for $z_+$, which is solved numerically, from 
which the other downstream quantities can be deduced. The first step in obtaining the equation for
$z_+$ involves the invariance across the shock of $na$, $V^\alpha V_\alpha$, $W^\alpha V_\alpha$, 
$W^\alpha l_\alpha-V^\alpha V_\alpha$ and
$X^\alpha X_\alpha$ where $X^\alpha=W^\alpha-(a^2nG+n/z+|b|^2/2)l^\alpha$. This leads to five equations 
for the five unknowns $a_+$, $|b|_+^2$, $d_+$, $n_+$ and $z_+$. These are
\eqb
[na]&=&0
\label{manilejump1}\\
\left[a^2|b|^2-d^2\right]&=&0
\label{manilejump2}\\
\left[dG\right]&=&0
\label{manilejump3}\\
\left[a^2nG+{n\over z}+{|b|^2\over 2}\right]&=&0
\label{manilejump4}\\
\left[a^2n^2G^2\left(1+a^2\right)+{2n\over z}\left(d^2-a^2|b|^2\right)+2a^2|b|^2nG\right]&=&0
\label{manilejump5}
\eqe
For algebraic simplicity we introduce the quantities $M\equiv na$, $Q\equiv (a^2|b|^2-d^2)/M$,
$P\equiv dG/\sqrt{M}$ and $D_+\equiv(Q+P^2/G_+^2)/2$ and the five equations 
can be written as \cite{majoranaanile87}
\eqb
[M]=[Q]=[P]&=&0
\label{majumps}\\
a_+G_++{1\over a_+z_+}+{D_+\over a_+^2}&=&c_1
\label{azjump1}\\
(1+a_+^2)G_+^2+{2\over a_+}\left(2D_+G_+-{Q\over z_+}\right)&=&c_2
\label{azjump2}
\eqe
where $c_1$ and $c_2$ are determined by the given upstream state. Equations (\ref{azjump1}) and
(\ref{azjump2}) contain just the two unknowns $a_+$ and $z_+$ since $D_+$ 
and $G_+$ are functions 
of $z_+$ and all other quantities are given by the upstream parameters. These two jump conditions
can be manipulated into a linear equation for $a_+$
\eqb
a_+={{\left(c_2+G_+/z_+-G_+^2\right)\beta
-\left(\alpha+D_+G_+\right)\alpha c_1}\over
{c_1\beta G_+-\alpha^2G_+}}.
\label{apluslinear}
\eqe
where
\eqb
\alpha\equiv3D_+G_+-{2Q\over z_+}\;\;\;\;{\rm and}\;\;\;\;
\beta\equiv{4D_+G_+\over z_+}-{2Q\over z_+^2}-D_+G_+^2+D_+c_2.
\nonumber
\eqe
Inserting into either (\ref{azjump1}) or (\ref{azjump2}) gives a transcendental equation for $z_+$ 
which can be solved by a root finding algorithm.

For a dynamically unimportant magnetic field the flow velocities are directed along the shock
normal so that Eq.(\ref{bparformula}) holds. A second relation between upstream and downstream 
magnetic fields can be obtained from (\ref{manilejump2}) which becomes
\eqb
\Gamma_-^2B_-^2\left(v_-^2-\cos^2\Phi_-\right)=
\Gamma_+^2B_+^2\left(v_+^2-\cos^2\Phi_+\right).
\nonumber
\eqe
When combined with Eq.(\ref{bparformula}) this gives the relationship (Eq.\ref{khformula}) between 
$B_+$, the upstream parameters and the shock compression ratio which is obtained from the hydrodynamic
jump conditions.


\section{The eigenvalues of the relativistic transport equation}
\label{appendixqj}

In the diffusion approximation, it is assumed that the distribution
function is everywhere close to isotropy. Applying this idea to
the region far upstream of a shock front suggests that
some information about the
eigenvalue $\Lambda^v_{i=1}$ and its eigenfunction $Q^v_{i=1}(\mu)$ for small $v$.
may be gained by a perturbative approach. This is indeed the case, and one finds
(assuming $D_{\mu\mu}$ is an even function of $\mu$)
\eqb
Q^v_1(\mu)
&=&1-{\Lambda^v_1\over2}\int_{-1}^{+1}\,d\mu'\,
{1-{\mu'}^2\over D_{\mu'\mu'}}.
\label{eq4.13a?}
\eqe
and
\eqb
\Lambda^v_1&=&8v\left[\int_{-1}^{+1}\,d\mu\,
{\left(1-\mu^2\right)^2\over D_{\mu\mu}}\right]^{-1}.
\label{eq4.16?}
\eqe
This can be expressed in terms of the spatial diffusion coefficient $\kappa$, which relates the gradient of the 
particle density to the flux:
\eqb
\kappa&=&c^2/\Lambda^v_1
\eqe
(where we have explicitly reintroduced the particle speed $c$) in agreement with the expression given, for 
example, by Skilling \cite{skilling75}).

In the ultrarelativistic case, particles which cross the shock from downstream
to upstream have their pitch angles 
concentrated in a very narrow cone $-1<\mu<v_-$, as seen in the upstream rest
frame. This indicates that a change of independent 
variable is appropriate. Choosing $y=(1+\mu)/(1-v)$, and approximating
$D_{\mu\mu}$ by the first term in its Taylor expansion about $\mu=-1$:
$D_{\mu\mu}\approx d(1+\mu)$ transforms Eq. (\ref{eq4.2?}) into 
\eqb
yQ'' + Q' + \epsilon^2\Lambda (y-1) Q/d&=&0
\label{ultrarel}
\eqe
(\cite{kirkschneider89})
where a prime indicates differentiation with respect to $y$ and we have
simplified the notation for $Q$ and $\Lambda$. 
We have also introduced the small parameter 
$\epsilon=1-v\approx 1/(2\Gamma^2)$. For large $y$, the solutions to this
equation have the asymptotic dependence $Q\sim{\rm
exp}(\pm\epsilon y\sqrt{\Lambda/d})$, whereas the regular solution close to the
singular point $y=0$ has the dependence $Q\propto 1-(\Lambda\epsilon^2/3d)y+{\rm
O}(y^2)$. Therefore, bounded solutions are of the form
\eqb
Q(y)&=&{\rm exp}(-\epsilon y\sqrt{\Lambda/d})\sum_{n=0}^N a_n y^n ,
\eqe
i.e., a polynomial with an exponential factor. Inserting this into
Eq.~(\ref{ultrarel}) and requiring that $a_{N+1}$ vanish yields the positive
eigenvalues:
\eqb
\Lambda_i^v\approx d(2i-1)^2/\epsilon^2.
\eqe
In addition, the following recursion relation is obtained:
\eqb
a_{n+1}&=&-{2(2N+1)(N-n)\over(n+1)^2}\,a_n.
\eqe
The eigenfunctions $Q_i^v(\mu)$ are found from these equations by setting $N=i-1$ and $y=(1+\mu)/(1-v)$.



\section*{References}
\begin{thebibliography}{10}


\bibitem{gaidosetal96}
Gaidos J.A. et al. 1996 \nature{383}{318}{}


\bibitem{metzgeretal97}
Metzger, M.R., et al. 1997 \nature{387}{87}{}


\bibitem{rees66}
Rees M.J. 1966 \nature{211}{468}{470}

\bibitem{vermeulencohen94}
Vermeulen, R.C., Cohen, M.H. 1994 \apj{430}{467}{}

\bibitem{taub}
Taub, A.H. 1948 \physrev{74}{328}{}

\bibitem{dehoffmannteller50}
de~Hoffmann F., Teller E. 1950 \physrev{80}{692}{}

\bibitem{akhiezerpolovin59}
Akhiezer I.A., Polovin R.V. 1959 \sovphysjetp{9}{1316}{1320}

\bibitem{lichnerowicz70}
Lichnerowicz A., 1970 \physicascripta{2}{221}{225}

\bibitem{webbetal87}
Webb G.M., Zank G.P., McKenzie J.F. 1987 \jplasmaphys{37}{117}{}

\bibitem{majoranaanile87}
Majorana A., Anile A.M., 1987 \physfluids{30}{3045}{3049}

\bibitem{applcamenzind88}
Appl S., Camenzind M. 1988 \aanda{206}{258}{}

\bibitem{kundtkrotscheck82}
Kundt W., Krotscheck E., 1982 \aanda{83}{1}{}

\bibitem{landaulifshitz59}
Landau, L. D. and Lifshitz, E. M. 1959
{\sl Fluid Mechanics} (London: Pergamon Press).

\bibitem{blandfordmckee76}{Blandford, R. D. and McKee, C. F.
1976 {\sl Phys. Fluids} {\bf 19}, 1130.}

\bibitem{koenigl80}
K\"onigl, A. 1980 \physfluids{23}{1083}{}

\bibitem{akhiezeretal59}
Akhiezer A.I., Liubarskii G.I., Polovin R.V. 1959 \sovphysjetp{8}{507}{511}

\bibitem{kenneletal83}
Kennel C.F., Edmiston J.P., Hada T. 1983
in \lq Collisionless Shocks in the Heliosphere: A Tutorial Review\rq\ 
Eds: B.T. Tsurutani, R.G. Stone, Geophysical Monograph~34, American Geophysical
Union, Washington D.C..


\bibitem{vanputten91}
van~Putten M.H.P.M. 1991 \commmathphys{141}{63}{72}

\bibitem{vanputten93}
van~Putten M.H.P.M. 1991 \jcompphys{105}{339}{353}

\bibitem{falleetal98}
Falle S.A.E.G., Komissarov S., Joarder P. 1998 \mnras{297}{265}{277}

\bibitem{isenberg86}
Isenberg P.A., 1986 \jgr{91}{1699}{1700}{}

\bibitem{kirkheavens89}
Kirk, J.G., Heavens, A.F. 1989 \mnras{239}{995}{1011}

\bibitem{begelmankirk90}
Begelman, M.C., Kirk, J.G. 1990 \apj{353}{66}{80}

\bibitem{lichnerowicz67}
Lichnerowicz A., 1967 \lq Relativistic Hydrodynamics
and Magnetohydrodynamics\rq, (Benjamin, New York) page 172.

\bibitem{reesgunn74}
Rees M.J., Gunn J.E. 1974 \mnras{167}{1}{}

\bibitem{kennelcoroniti84}
Kennel C.F., Coroniti F.V., 1984 \apj{283}{694}{709}

\bibitem{michel91}
Michel, F.C. 1991, \lq Theory of Neutron Star Magnetospheres\rq, University of
Chicago Press, Chicago.

\bibitem{ballardheavens91}
Ballard K.R., Heavens A.F., 1991 \mnras{251}{438}{448}

\bibitem{hallsturrock67}
Hall, D.E., Sturrock, P.A. 1967 \physfluids{10}{2620}{}

\bibitem{melrose69}
Melrose, D.B. 1969 \ass{4}{143}{}

\bibitem{luhmann76}
Luhmann, J.G. 1976 \jgr{81}{2089}{}

\bibitem{achatzetal91}
Achatz, U., Steinacker, J., Schlickeiser, R. 1991 \aanda{250}{266}{}

\bibitem{lee82}
Lee M.A. 1982 \jgr{87}{5063}{}

\bibitem{lee83}
Lee M.A. 1983 \jgr{88}{6109}{}

\bibitem{jokipii66}
Jokipii J.R., 1966 \apj{146}{480}{}

\bibitem{hasselmannwibberenz70}
Hasselmann K., Wibberenz G., 1970 \apj{162}{1049}{}

\bibitem{parker65}
Parker, E.N. 1965 \planetspacesci{13}{9}{}

\bibitem{gleesonaxford67}
Gleeson, L.J., Axford, W.I. 1967 \apjletts{149}{L115}{}

\bibitem{krymsky77}
Krymsky, G.F. 1977 \sovphysdokl{22}{327}{328}

\bibitem{axfordetal77}
Axford, W.I., Leer, E., Skadron, G. 1977 \icrcplodiv{11}{132}{137}

\bibitem{bell78}
Bell, A.R. 1978 \mnras{182}{147}{}

\bibitem{blandfordostriker78}
Blandford, R.D., Ostriker, J.P. 1978 \apjletts{221}{L29}{L32}

\bibitem{drury83}
Drury, L.O'C. 1983 \repprogphys{46}{973}{1027}

\bibitem{kirketal94}
Kirk J.G., Melrose D.B., Priest E.R., 1994
{\it Plasma Astrophysics}, Saas Fee Advanced Course 24, Eds.: A.O.Benz, T.J.-L. Courvoisier, 
(Springer-Verlag, Berlin).

\bibitem{gallantetal98}
Gallant Y.A., Achterberg A., Kirk J.G. 1998
in \lq\lq Rayos c\'osmicos 98\rq\rq,
Proc.\ 16th.\ European Cosmic Ray Symposium, Alcal\`a
Ed.\ J.~Medina, Univ. de Alcal\`a (Madrid) p.~371

\bibitem{hoshinoetal92}
Hoshino M., Arons J., Gallant Y.A., Langdon A.B. 1992 \apj{390}{454}{}

\bibitem{gallantetal92}
Gallant Y.A., Hoshino M., Langdon A.B., Arons J., Max C.E., 1992 
\apj{391}{73}{101}

\bibitem{gallantarons94}
Gallant Y.A., Arons J. 1994 \apj{435}{230}{260}

\bibitem{scheuer89}
Scheuer P., 1989 in \lq Hot-Spots in Extragalactic Radio Sources\rq\ 
Eds.~K.~Meisenheimer, H.-J.~R\"oser, (Springer-Verlag, Berlin) 
Lecture Notes in Physics {\bf 327}, 159

\bibitem{parker58}
Parker, E.N. 1958 \physrev{109}{1328}{1344}

\bibitem{schatzman63}
Schatzman, E. 1963 \annalastro{26}{234}{239}

\bibitem{decker88}
Decker, R.B. 1988 \spscrev{48}{195}{262}

\bibitem{whippleetal86}
Whipple, E.C., Northrop, T.G., Birmingham, T.J. 1986
\jgr{91}{4149}{}

\bibitem{abramowitzstegun72}
Abramowitz, M., Stegun, I.A.  1972 {\it Handbook of
Mathematical Functions} (Washington DC: National Bureau of
Standards)

\bibitem{vanderlaan62}
van der Laan, H. 1962 \mnras{124}{125}{}

\bibitem{gieseleretal99}
Gieseler U.D.J:, Gallant Y.A., Kirk J.G., Achterberg A.
1999 A\&A, in press, (available as astro-ph/9902079)

\bibitem{peacock81}
Peacock, J.A. 1981 \mnras{196}{135}{}

\bibitem{kirkschneider87a}
Kirk J.G., Schneider P. 1987a \apj{315}{425}{433}

\bibitem{ostrowski91}
Ostrowski M., 1991 \mnras{249}{551}{559}

\bibitem{naitotakahara95}
Naito T., Takahara F., 1995 \mnras{275}{1077}{1092}

\bibitem{schneiderkirk89}
Schneider P., Kirk J.G., 1989 \aanda{217}{344}{350}

\bibitem{kirkschneider89}
Kirk, J.G., Schneider, P. 1989 \aanda{225}{559}{568}

\bibitem{malkovvoelk95}
Malkov M.A., V\"olk H.J., 1995 \aanda{300}{605}{}

\bibitem{kirkschneider88}
Kirk, J.G., Schneider, P. 1988 \aanda{201}{177}{188}

\bibitem{baringkirk91}
Baring M.G., Kirk J.G., 1991 \aanda{241}{329}{342}

\bibitem{druryvoelk81}
Drury L.O'C., V\"olk H.J. 1981 \apj{248}{344}{}

\bibitem{kirketal88}
Kirk, J.G., Schlickeiser, R., Schneider, P. 
1988 \apj{328}{269}{274}

\bibitem{ince56}
Ince E.L., 1956 \lq Ordinary Differential Equations\rq\ (Dover, New York)

\bibitem{kirk88}
Kirk J.G., 1988 Habilitationsschrift, LMU M\"unchen

\bibitem{heavensdrury88}
Heavens, A.F., Drury, L.O'C. 1988 \mnras{235}{997}{}

\bibitem{voelketal74}
V\"olk, H.J., Morfill, G.,
Alpers, W., Lee, M.A. 1974 \ass{26}{403}{}

\bibitem{bednarzostrowski98}
Bednarz J., Ostrowski M., 1998 \physrevletts{80}{3911}{3914}

\bibitem{ballardheavens92}
Ballard K.R., Heavens A.F., 1992 \mnras{259}{89}{94}

\bibitem{ostrowski93}
Ostrowski M., 1993 \mnras{264}{248}{}

\bibitem{raxwhite92}
Rax J.M., White R.B., 1992 \physrevletts{68}{1523}{}

\bibitem{kirketal96}
Kirk J.G., Duffy P., Gallant Y.A., 1996 \aanda{314}{1010}{1016}

\bibitem{achterberg88}
Achterberg A., 1988 \mnras{232}{323}{337}

\bibitem{gardiner83}
Gardiner C.W., 1983 \lq Handbook of Stochastic Methods\rq\ (Springer-Verlag,
Berlin) 

\bibitem{kirkschneider87b}
Kirk, J.G., Schneider, P. 1987b \apj{322}{256}{265}

\bibitem{kruellsachterberg94}
Kr\"ulls W., Achterberg A. 1994 \aanda{286}{314}{}

\bibitem{zachary87}{Zachary, A. 1987
{\sl Resonant Alfv\'en Wave Instabilities Driven by
Streaming Fast Particles}, Ph.D. thesis, Lawrence Livermore
National Laboratory, University of California.}

\bibitem{jonesellison91}
Jones, F.C., Ellison, D.C. 1991 \spscrev{58}{259}{}

\bibitem{ellisonetal90}
Ellison D.C., Jones F.C., Reynolds S.P., 1990 \apj{360}{702}{714}

\bibitem{quenbylieu89}
Quenby J.J., Lieu R., 1989 \nature{342}{654}{656} 

\bibitem{lieuetal94}
Lieu R., Quenby J.J., Drolias B., Naidu K., 1994 \apj{421}{211}{218}

\bibitem{bednarzostrowski96}
Bednarz J., Ostrowski M., 1996 \mnras{283}{447}{456}

\bibitem{ellisonetal96}
Ellison D.C., Baring M.G., Jones F.C., 1996 \apj{473}{1029}{}

\bibitem{waxman97}
Waxman, E. 1997 \apj{485}{L5}{L8}

\bibitem{galamaetal98}
Galama T.J., Wijers R.A.M.F., Bremer M., Groot P.J.,
Strom R.G., De~Bruyn A.G., Kouveliotou C., Robinson C.R.,
Van~Paradijs J. 1998 \apjletts{500}{L101}{104}

\bibitem{marcowithetal95}
Marcowith A., Henri G., Pelletier, G. 1995 \mnras{277}{681}{}

\bibitem{romanovalovelace97}
Romanova M.M., Lovelace R.V.E. 1997 \apj{475}{97}{105}

\bibitem{levinson98}
Levinson A. 1998 \apj{507}{145}{154}

\bibitem{kirketal98}
Kirk J.G., Rieger F.M., Mastichiadis A. 1998 \aanda{333}{452}{458}

\bibitem{begelmanblandfordrees84}
Begelman, M.C., Blandford, R.D., Rees, M.J. 1984 \revmodphys{56}{255}

\bibitem{meisenheimerheavens86}
Meisenheimer, K., Heavens, A.F. 1986 \nature{323}{419}{422}

\bibitem{heavensmeisenheimer87}
Heavens, A.F., Meisenheimer, K. 1987 \mnras{225}{335}{353}

\bibitem{meisenheimerroser89}
Meisenheimer, K., R\"oser, H.-J., Hiltner, P.R., Yates, M.G., Longair, M.S.,
Chinis, R., Perley, R.A. 1989 \aanda{219}{63}{}

\bibitem{wilsonscheuer83}
Wilson, M.J., Scheuer, P.A.G. 1983 \mnras{205}{449}{463}

\bibitem{gomezetal97}
Gomez J.L., Marti J.M., Marscher A.P., Iba\~nez J.M., Alberdi A. 1997
\apjletts{482}{L33}{} 

\bibitem{mioduszewskietal97}
Mioduszewski A.J., Hughes P.A., Duncan G.C. 1997 \apj{476}{649}{}

\bibitem{komissarovfalle97}
Komissarov S., Falle S.A.E.G. 1997 \mnras{288}{833}{848}

\bibitem{matthewsscheuer90a}
Matthews, A.P., Scheuer, P.A.G. 1990 \mnras{242}{616}{}

\bibitem{matthewsscheuer90b}
Matthews, A.P., Scheuer, P.A.G. 1990 \mnras{242}{623}{}

\bibitem{marschergear85}
Marscher A.P., Gear, W.K. 1985 \apj{298}{114}{}

\bibitem{hughesalleraller91}
Hughes P.A., Aller H.D., Aller M.F. 1991 \apj{374}{57}{}

\bibitem{marschertravis96}
Marscher A.P., Travis J.P., 1996 \aandasuppl{120C}{537}{} 

\bibitem{wagnerwitzel95}
Wagner S.J., Witzel A. 1995 \araa{33}{163}{}

\bibitem{aharonianetal99}
Aharonian F., et al. 1999 \aanda{342}{69}{}

\bibitem{dermerschlickeiser93}
Dermer C.D., Schlickeiser R. 1993 \apj{416}{458}{}

\bibitem{ghisellinimaraschidondi96}
Ghisellini G., Maraschi L., Dondi L. 1996 \aandasuppl{120C}{503}{}

\bibitem{steckerdejagersalamon96}
Stecker F.W., de Jager O.C., Salamon M.H. 1996 \apjletts{473}{L75}{}

\bibitem{mastichiadiskirk97}
Mastichiadis A., Kirk, J.G. 1997 \aanda{320}{19}{}

\bibitem{mannheimwesterhoff96}
Mannheim K., Westerhoff S., Meyer H., Fink H.-H. 1996 \aanda{315}{77}{}

\bibitem{cavallorees78}
Cavallo, G., Rees, M.J. 1978 \mnras{183}{359}{}

\bibitem{goodman86}
Goodman, J. 1986 \apjletts{308}{L47}{}

\bibitem{paczynski86}
Paczynski, B. 1986 \apjletts{308}{L43}{}

\bibitem{reesmeszaros92}
Rees, M.J., M\'esz\'aros, P. 1992 \mnras{258}{41P}{}

\bibitem{blandfordmckee77}
Blandford R.D., McKee C.F. 1977 \mnras{180}{343}{}

\bibitem{meszaroslagunarees93}
M\'esz\'aros, P., Laguna, P., Rees, M.J. 1993 \apj{415}{181}{}

\bibitem{piransheminarayan93}
Piran, T., Shemi, A., Narayan, R. 1993 \mnras{263}{861}{}

\bibitem{saripiran95}
Sari, R., Piran, T. 1995 \apjletts{455}{L143}{}

\bibitem{reesmeszaros94}
Rees, M.J., M\'esz\'aros, P. 1994 \apjletts{430}{L93}{}

\bibitem{saripiran97}
Sari, R., Piran, T. 1997 \mnras{287}{110}{}

\bibitem{vietri95}
Vietri, M. 1995 \apj{453}{883}{}

\bibitem{waxman95}
Waxman, E. 1995 \physrevletts{75}{386}{}

\bibitem{meszarosrees93}
M\'esz\'aros, P., Rees, M.J. 1993 \apjletts{418}{L59}{}

\bibitem{dermer98}
Dermer, C.D. 1998 \apjletts{501}{157}{}

\bibitem{wijersreesmeszaros97}
Wijers, R.A.M.J., Rees, M.J., M\'esz\'aros, P. 1997 \mnras{288}{L51}{}

\bibitem{synge57}
Synge, J. L. 1957 {\sl The Relativistic Gas}
(Amsterdam: North-Holland Publishing Co.)

\bibitem{johnsonmckee71}
Johnson M.H., McKee C.F. 1971 \physrev{D3}{858}{}

\bibitem{skilling75}
Skilling, J. 1975 \mnras{172}{557}{}

\end{thebibliography}
\end{document}